\documentclass[numberedappendix]{emulateapj}
\usepackage{natbib}
\usepackage{graphicx}
\usepackage{dcolumn}
\usepackage{bm}
\usepackage{amsmath}
\usepackage{amstext}
\usepackage{pstricks}
\usepackage{color}
\usepackage{hyperref}
\renewcommand{\baselinestretch}{1.18}

\newcommand\refsec[1]{\S \ref{sec:#1}}

\newcommand{\degree}{^\circ}

\newcommand{\gps}{\ensuremath{g_{\rm P1}}}
\newcommand{\rps}{\ensuremath{r_{\rm P1}}}
\newcommand{\ips}{\ensuremath{i_{\rm P1}}}
\newcommand{\zps}{\ensuremath{z_{\rm P1}}}
\newcommand{\yps}{\ensuremath{y_{\rm P1}}}

\newcommand{\grizy}{\gps\rps\ips\zps\yps}
\newcommand{\PS}{\protect \hbox {Pan-STARRS1}}

\begin{document}

\title{Hypercalibration: \\A Pan-STARRS1-based recalibration of the Sloan Digital Sky Survey
Photometry}

\author{Douglas P. Finkbeiner\altaffilmark{1,2}, 
  Edward F. Schlafly\altaffilmark{3}, 
  David J. Schlegel\altaffilmark{4},
  Nikhil Padmanabhan\altaffilmark{5}, 
  Mario {Juri{\'c}}\altaffilmark{6}, 
  William S. Burgett\altaffilmark{7}, 
  Kenneth C. Chambers\altaffilmark{7}, 
  Larry Denneau\altaffilmark{7}, 
  Peter W. Draper\altaffilmark{8}, 
  Heather Flewelling\altaffilmark{7}, 
  Klaus W. Hodapp\altaffilmark{7}, 
  Nick Kaiser\altaffilmark{7}, 
  E. A. Magnier\altaffilmark{7}, 
  N. Metcalfe\altaffilmark{8},
  Jeffrey S. Morgan\altaffilmark{7}, 
  Paul A. Price\altaffilmark{9}, 
  Christopher W. Stubbs\altaffilmark{2},
  John L. Tonry\altaffilmark{7}}

\altaffiltext{1}{Institute for Theory and Computation,
  Harvard-Smithsonian Center for Astrophysics, 
  60 Garden Street, MS-51, Cambridge, MA 02138 USA} 

\altaffiltext{2}{Department of Physics, 
  Harvard University, Cambridge, MA 02138 USA}

\altaffiltext{3}{Max-Planck-Institut f\"ur Astronomie, K\"onigstuhl 17,
  D-69117 Heidelberg, Germany}

\altaffiltext{4}{Lawrence Berkeley National Lab, 1 Cyclotron Rd, Berkeley CA 94720, USA}

\altaffiltext{5}{Department of Physics, Yale University, 260 Whitney Ave, New
  Haven, CT 06520, USA}

\altaffiltext{6}{LSST Corporation, 933 N. Cherry Avenue, Tucson, AZ 85721, USA}

\altaffiltext{7}{Institute for Astronomy, University of Hawaii at Manoa, Honolulu, HI 96822, USA}

\altaffiltext{8}{Department of Physics, Durham University, South Road, Durham DH1 3LE, UK}

\altaffiltext{9}{Department of Astrophysical Sciences, Princeton University, Princeton, NJ 08544, USA}

%{The Spatial Uniformity of the Sloan Digital Sky Survey Photometric Calibration}
%\author[Finkbeiner et al]{
%Douglas P. Finkbeiner$^{1}$, 
%Nikhil Padmanabhan$^{2}$ \\
%$^{1}$ Harvard-Smithsonian Center for Astrophysics, Harvard University, 60 Garden St., Cambridge, MA 02138 \\
%$^{2}$ Dept. of Physics, Yale University, 260 Whitney Ave, New Haven, CT 06520 \\ 

\begin{abstract}
  We present a recalibration of the Sloan Digital Sky Survey (SDSS)
  photometry with new flat fields and zero points derived from
  Pan-STARRS1 (\PS).  Using PSF photometry of 60 million stars with
  $16 < r < 20$, we derive a model of amplifier gain and flat-field
  corrections with per-run RMS residuals of 3 millimagnitudes (mmag)
  in $griz$ bands and 15 mmag in $u$ band.  The new photometric zero
  points are adjusted to leave the median in the Galactic North
  unchanged for compatibility with previous SDSS work.  We also
  identify transient non-photometric periods in SDSS (``contrails'')
  based on photometric deviations co-temporal in SDSS bands.  The
  recalibrated stellar PSF photometry of SDSS and PS1 has an RMS
  difference of \{9,7,7,8\} mmag in $griz$, respectively, when
  averaged over $15'$ regions.
\end{abstract}

\section{Introduction}
\label{sec:intro}
One of the challenges of wide-field surveys is ensuring the uniformity of
photometric calibration over the survey area.  
The traditional approach of calibrating to networks of standard stars requires
a transfer of calibrations between systems that would require perfect knowledge
of both the bandpasses and the stellar SEDs to avoid introducing systematic
errors.  Furthermore, standard stars are typically too bright to be observed
accurately under survey conditions, introducing additional steps in the
calibration process. 

% ====================== Figure =======================================

\begin{figure*}
\begin{center}
\includegraphics[width=0.49\textwidth]{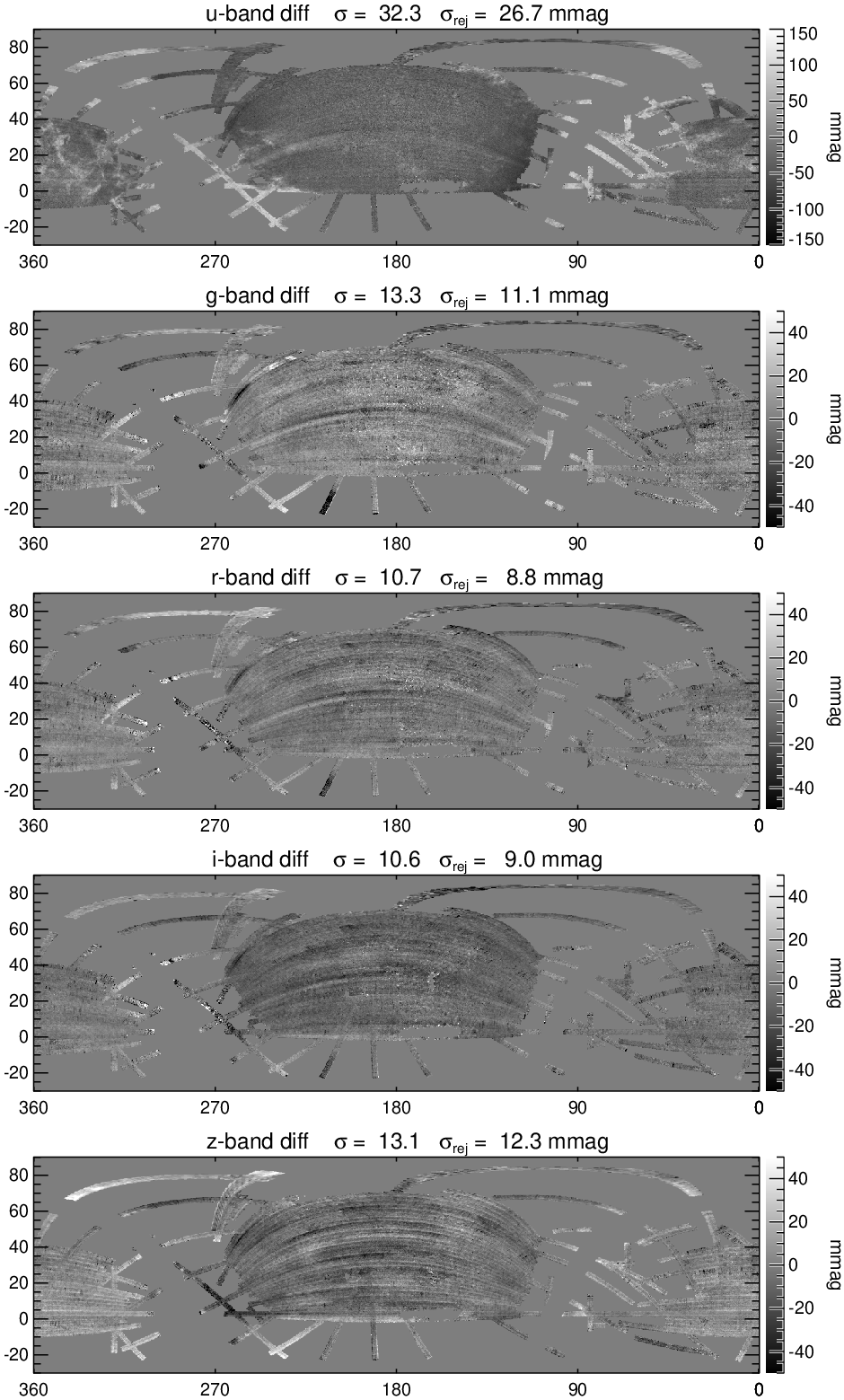}
\includegraphics[width=0.49\textwidth]{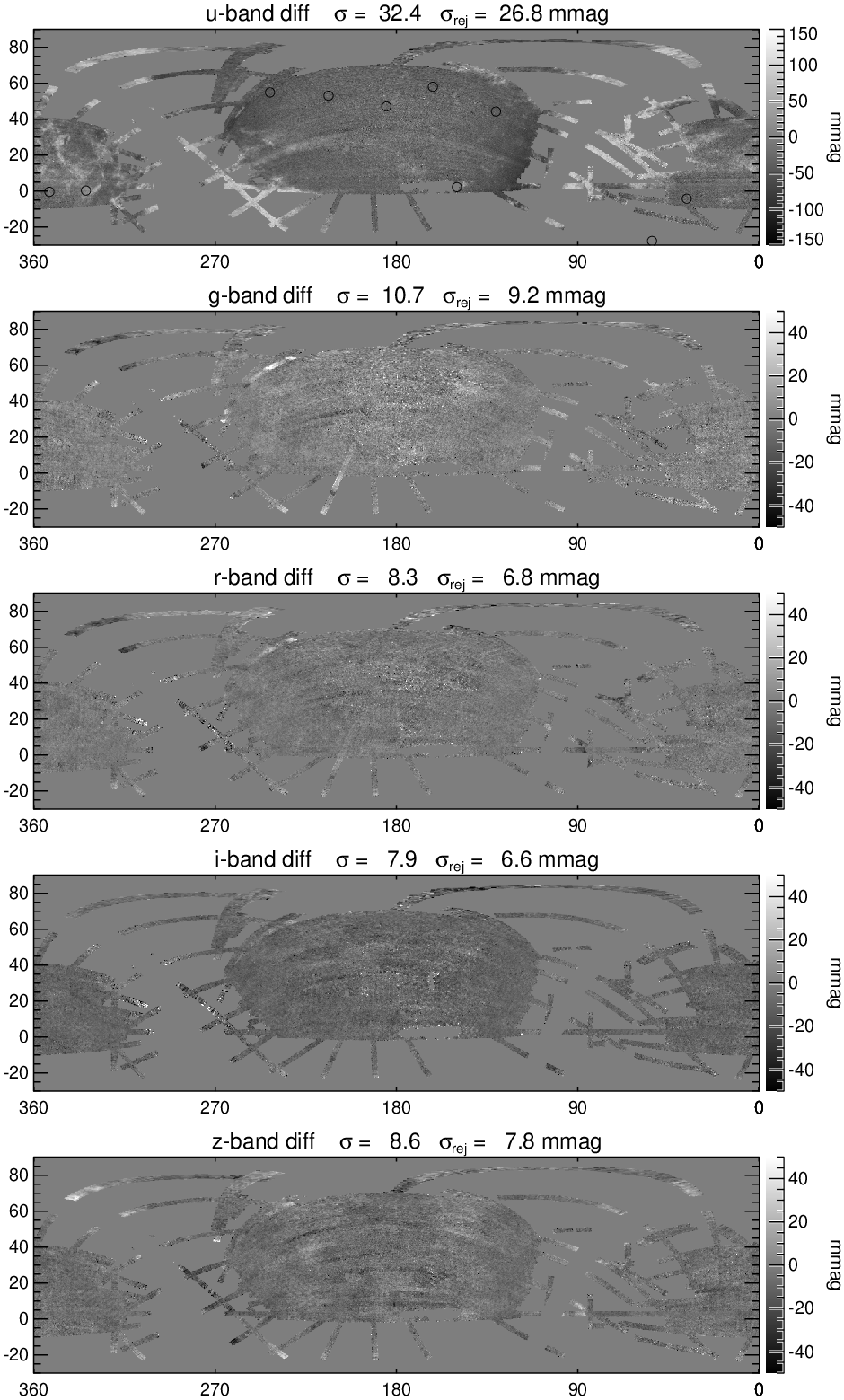}
\caption{\emph{Left panels:} the mean difference (PS1 minus SDSS) in 15
  arcminute pixels using the color transformations defined by
  \cite{standards:2015} (see \refsec{colortrans}).  In dark regions,
  SDSS underestimates the stellar fluxes.  Stripes corresponding to
  the SDSS scan pattern are readily apparent, as well as offsets
  between Galactic North and South.  The $u$-band difference map
  involves an extrapolation from $\gps$ and is far more sensitive to
  metallicity and extinction than the other bands.  The RMS of the
  difference per $15'$ pixel is given, with and without $3\sigma$
  outlier rejection.  Note the increased grayscale range in $u$-band.
  \emph{Right panels:} Same as left panels, but after the
  recalibration described in \S\ref{sec:calib}.  Striping is reduced,
  and the North-South offsets removed.  The calibration is adjusted to
  preserve the median magnitude at $b > 20\degree$ so as to minimally
  perturb SDSS results in the northern Galactic cap (see \S
  \ref{sec:calib}).  Dust cirrus is visible in the top panels, as it
  affects the extrapolation from PS1 bands to $u$ band more than the
  other bands.  Because of this, and possible gradients due to
  metallicity variations, we do not apply zero-point corrections to
  $u$ band.  Positions of the 10 Medium-Deep Fields are indicated by
  black circles in the upper right panel.  }
\label{fig:healdiff5}
\end{center}
\end{figure*}

The Sloan Digital Sky Survey \citep[SDSS;][]{York:2000} took a different approach to
the calibration problem, now generally referred to as ``ubercalibration''
\citep{ubercal:2008}. The underlying idea is simple: the flux from a star is
assumed to be constant\footnote{We ignore variability, which can be mitigated
  by averaging multiple stars in a given region with appropriate outlier
  rejection.}, and therefore comparisons of instrumental flux in repeat
observations can be used to infer calibration parameters.  By construction,
such comparisons are all done on a consistent photometric system and using
comparable observations, obviating the need for bandpass transformations
between the standard system and the survey system.  Indeed, nearly every
multiply observed star in the survey plays the role of a ``standard'' star
without detailed knowledge of its SED.
Such an algorithm is based on differences in magnitude and can only determine
relative calibrations up to an overall offset, cleanly separating the problem
of relative calibration from that of absolute calibration to e.g. AB
magnitudes \citep{Oke:1983} or physical units.

This calibration scheme requires a network of overlapping observations
that connect the entire survey area.  The SDSS camera is found to have
a nearly stable photometric zero-point during a night, with small
drifts in the atmospheric extinction parameterized by a time
derivative term ($\dot{k}$).  This stability allows widely
separated regions of sky to be connected by a single hours-long drift
scan.  This spatial and temporal structure of overlapping observations
ultimately determines the complexity of the calibration model and
therefore the photometric accuracy available. \cite{ubercal:2008}
present a detailed discussion of
the various degeneracies possible due to insufficient overlaps. Since the SDSS
was not designed with ubercalibration in mind, imaging overlaps are limited to
overlaps of interleaved ``strips'', and the survey poles where the imaging
great circles converge.  A set of fast scans with the SDSS camera, binned
$4\times4$, were obtained from May 2002 to April 2004 that cross the main
survey scans nearly perpendicularly, but the bulk of these photometric data
span only 23h $< $RA$ < $8h.  
The addition of 3,000 square degrees of imaging in
the South Galactic Cap, released as part of SDSS-III Data Release 8 \citep{DR8},
is only weakly connected to the main SDSS-I/II observations in the North
Galactic Cap.  The low number of drift scans connecting the Galactic
south with the Galactic north results in uncertainty in their relative
calibration. 

At the time the SDSS ubercalibration was released, it was not possible
to directly determine the achieved spatial uniformity of the
calibration. Using simulations, \cite{ubercal:2008} estimated the
lower bound on these errors to be $\sim$ 8 millimagnitudes (hereafter
mmag) in the $g$, $r$, $i$, and $z$ bands and $\sim$ 13
mmag in the $u$ band. Consistent estimates of the calibration errors
have been obtained by \cite{Schlafly:2010} using the blue tip of the stellar
locus, but the direct interpretation of these results as calibration errors is
complicated by spatial variations in the properties of stellar populations,
e.g. metallicity.  This paper presents an astrophysics-free estimate of the
spatial uniformity of the SDSS calibration by direct comparison with data from
the \PS\ $3\pi$ survey. 
\PS\ has the great advantage that it observed the survey footprint 6-8
times in each filter, providing high redundancy.  Furthermore, ten
`Medium-Deep' fields \citep[MD01-MD10 in][]{Tonry:2012md} were
observed hundreds of times per filter with longer exposures, resulting
in a photometric solution far more rigid than that of SDSS.  See
upper-right panel of Figure \ref{fig:healdiff5} for MD field locations. 

We find that the SDSS achieves the claimed photometric stability of
(20,10,10,10,20) mmag in (\emph{ugriz}) in the North Galactic Cap, but
contains spatially coherent offsets on the scale of fields, runs, and
even hemispheres (Figure \ref{fig:healdiff5}), as well as a small
fraction of significant non-Gaussian outliers.  In the following, we
present a new determination of flat fields and per-run offsets based
on the comparison to PS1.  We emphasize that it is the rigidity of the
PS1 solution on large angular scales, \emph{not} a superior
per-exposure stability, that makes it an excellent foundation upon
which to build an improved calibration of SDSS.  Indeed, neither the
statistical uncertainties nor systematic errors on small scales in PS1 are very much
better than SDSS, but they are \emph{different}.

The SDSS calibration also fails to capture short periods of
non-photometricity caused by small clouds or contrails.  We identify
sudden (in SDSS observation time) deviations in the PS1-SDSS
difference and record them in two new \texttt{CALIB\_STATUS} mask
bits.

In \S 2 we present the SDSS and PS1 data, the sample of stars, and
color transformations between the systems.  The new calibration is
presented in \S 3, and detection of unphotometric periods of time is
described in \S 4.

\section{Data}
\label{sec:data}
\subsection{SDSS} 
The Sloan Digital Sky Survey (SDSS) has been in operation since 1998, and is
now in its third phase (SDSS-III).  It uses a dedicated 2.5-m telescope
\citep{Gunn:2006} at Apache Point Observatory in New Mexico to perform a
variety of surveys.  In this work, we use photometry from a 30 CCD camera with
1.5 deg$^2$ effective field of view \citep{Gunn:1998} that imaged 14,555
deg$^2$ (about 35\%) of the sky in 5 broad bands (\emph{ugriz})
\citep{Fukugita:1996} between Sep 1998 and Nov 2009, after which the camera
was retired from operation.  The CCDs are arranged with one chip per band in
each of 6 \emph{camera columns}, and operate in a drift-scan mode such that
objects pass over the 5 filters in a period of 5.4 minutes.  A contiguous
period of drift scan is a \emph{run} and may last for up to 10 hours.  The
region imaged by the 6 camera columns is a \emph{strip} with 6 regions $13.5'$
wide separated by gaps of $12.5'$.  A subsequent strip fills in these gaps, and
together the two strips constitute a \emph{stripe} $2.5\degree$ wide.  Objects are
detected and characterized by a photometric pipeline \citep{Lupton:2001} and
astrometric and photometric calibrations are applied
\citep{Pier:2003,Ivezic:2004,Tucker:2006,ubercal:2008}.  As of Data Release 8
\citep[DR8;][]{DR8}, the imaging survey was completed.
We use DR9 photometry \citep{DR9}, which is identical to DR8 except for its
astrometric calibration tying the full survey to UCAC 2.0 \citep{UCAC2:2004}(details at
\url{http://www.sdss3.org/dr9}).  To emphasize that the photometry is
identical for the two data releases, we refer to them hereafter as DR8/DR9.
We use the \texttt{calibObj} files,\footnote{schema at
  \url{http://data.sdss3.org/datamodel/files}} trimmed versions of the
\texttt{photoObj} files containing the most commonly used parameters, 
and separated into star and galaxy files.  There is one 
stellar \texttt{calibObj} file for each run+camcol.  DR8/DR9 contains 764 runs.

We reject objects with flag bits 2,11,18,22, and 43 set, corresponding to
\texttt{EDGE}, \texttt{DEBLEND\_TOO\_MANY\_PEAKS}, \texttt{SATUR},
\texttt{BADSKY}, and \texttt{SATUR\_CENTER}, respectively.  Because of many
levels of outlier rejection, our results are
not sensitive to these choices. 

\begin{table*}
\begin{center}
\begin{tabular}{|r|r|l||}
\hline
CALIB\_STATUS bit name & Bit & Description \\
\hline
PHOTOMETRIC            &  0    &  Photometric observations  \\
UNPHOT\_OVERLAP        &  1    &  Unphotometric observations, calibrated based on overlaps with clear, ubercalibrated data;\\
                       &       &  ~~~~~done on a field-by-field basis. \emph{Use with caution.}  \\
UNPHOT\_EXTRAP\_CLEAR  &  2    &  Extrapolate the solution from the clear part of a night (that was ubercalibrated) \\
                       &       &  ~~~~~to the cloudy part. \emph{Not recommended for use.}  \\
UNPHOT\_EXTRAP\_CLOUDY &  3    &  Extrapolate the solution from a cloudy part of the night (where there is overlap) \\
                       &       &  ~~~~~to a region of no overlap. \emph{Not recommended for use.}  \\
UNPHOT\_DISJOINT       &  4    &  Data is disjoint from the rest of the survey.  Even though conditions may be photometric,\\
                       &       &  ~~~~~the calibration is suspect. \emph{Not recommended for use.}  \\
INCREMENT\_CALIB       &  5    &  Incrementally calibrated by considering overlaps with ubercalibrated data  \\
{\bf PS1\_UNPHOT }     &\bf{6} &  \bf{PS1 comparison reveals unphotometric conditions}  \\
{\bf PS1\_CONTRAIL}    &\bf{7} &  \bf{PS1 comparison shows slightly unphotometric conditions}  \\ 
PT\_CLEAR              &  8    &  (INTERNAL, DR8 and later) PT calibration for clear data  \\
PT\_CLOUDY             &  9    &  (INTERNAL, DR8 and later) PT calibration for cloudy data  \\
DEFAULT                &  10   &  (INTERNAL, DR8 and later) a default calibration used  \\
NO\_UBERCAL            &  11   &  (INTERNAL, DR8 and later) not uber-calibrated \\
INTERNAL               &  12   &  (INTERNAL USE) \\
{\bf PS1\_PCOMP\_MODEL}&\bf{13}&  \bf{PS1 Used PCA model for flats}  \\
{\bf PS1\_LOW\_RMS}    &\bf{14}&  \bf{PS1 comparison to SDSS has low noise}  \\
\hline
\end{tabular}
\end{center}
\caption{Calib\_status bits.  New ones are bold. 
}
\label{tbl:calib_status}
\end{table*}

\subsection{Pan-STARRS1}
The Pan-STARRS1 (\PS) $3\pi$ survey \citep{Kaiser:2010} and (Chambers et al., in
preparation) is a systematic imaging survey of $3/4$ of the sky north
of $\delta = -30\degree$ in five optical and near-infrared photometric
bands \citep[\grizy;][]{Tonry:2012}.  The survey is conducted with a
1.4 billion pixel, $3.3\degree$ field-of-view camera
\citep{Onaka:2008,Tonry:2009} on a dedicated 1.8m telescope
\citep{Hodapp:2004} located on Haleakala, Hawaii.  Any location in the
survey is observed repeatedly for a planned four times per year per
filter, conditions permitting, with exposure times of 43/40/45/30/30
seconds in the \gps/\rps/\ips/\zps/\yps-bands, respectively
\citep{Metcalfe:2013}. The median FWHM values in these bands are
1.27/1.16/1.11/1.06/1.01 arcseconds. Images are automatically
processed through the Image Processing Pipeline
\citep{Magnier:2006,Magnier:2007,Magnier:2008} to produce the object
catalog.  The data set used for this work includes three consecutive
seasons of observing, yielding up to twelve exposures per filter. Chip
and cell gaps, variable observing conditions, and technical problems
cause the survey depth to vary from place to place.  For point
sources, $5\sigma$ limits for the $3\pi$ survey (single exposure)
are 22.2, 22.2, 22.0, 21.2, 20.1 in $griz$\yps, respectively.  For
comparison, SDSS has stellar $5\sigma$ depth of 22.2, 23.1, 22.7,
22.2, 20.7 in $ugriz$.

The PS1 focal plane has 60 OTA \emph{chips}, each of which is an
$8\times8$ grid of independently addressable \emph{cells}.  The
\PS\ $3\pi$ survey covers the entire SDSS footprint, to similar depth
in similar filters (except u-band), allowing a straightforward
comparison between the two surveys after modest color transformations
\refsec{colortrans}.

The \PS\ photometric calibration \citep{Schlafly:2012} minimizes the
variance of repeat measurements of stars in much the same way as the SDSS
ubercalibration.  Schlafly et al.\ fit for a flat field (in
$2\times2$ cell regions), as well as a zero point and atmospheric extinction
term per night.  On nearly every night, PS1 observes a few of the 10 medium
deep (MD) fields, which have been observed hundreds of times per filter on dozens of
nights.  These serve as \emph{de facto} standard star fields in the calibration,
providing a rigid foundation on which to build the photometric solution for
the entire $3\pi$ survey.  Because of this, and the multiple coverings of the
$3\pi$ area, the PS1 solution is more rigid on large angular scales, tying
the northern and southern Galactic hemispheres together much better
than SDSS.  As an estimate of the photometric stability of this solution, the
zero points of repeat visits to the MD fields vary by less than 5 mmag in $gri$\zps
\citep[][\S3.1]{Schlafly:2012}.  On small (sub-degree) scales, the PS1 photometry may
not be more stable than SDSS, but the two surveys have uncorrelated systematic
errors, allowing precise derivation of SDSS photometric parameters by
comparison to PS1.

\subsection{Matching Catalogs}
\label{sec:match}

We match PS1 objects to SDSS detections of point sources (\texttt{objc\_type} = 6)
with $16 <r_{SDSS} <20$, using a match radius of $1''$.  
This is not a list of unique stars; stars observed multiple
times by SDSS are multiply counted in the following.  Of 118,582,000 matches, we select
111,980,000 that
pass the cuts on flags described above.  We discard stars within $15\degree$ of
the Galactic plane because of difficulty with high stellar density and with
color transformations in regions of high dust reddening, leaving 81,068,000
stars, or 72.4\% of the selected stars.

Approximately 11\% of the SDSS sample is marked unphotometric
(by \texttt{CALIB\_STATUS} bit 0, see Table \ref{tbl:calib_status}
and the SDSS-III web site\footnote{\href{http://www.sdss3.org/dr8/algorithms/bitmask_calib_status.php}{\texttt{\scriptsize http://www.sdss3.org/dr8/algorithms/bitmask\_calib\_status.php}}}), and
we exclude these stars from determination
of the flat fields. 
In terms of sky area, the fraction of photometric fields containing stars with $|b| >
15\degree$ is 87.5\% (813912/929827).  This corresponds to 27451 deg$^2$ of 31360
deg$^2$, much larger than the actual survey footprint because of run
overlaps and repeated scans of the equatorial stripe. 

When deriving new photometric offsets for runs, we require at least $3\degree$ of
the scan to be photometric and at $|b| > 15\degree$.
If these criteria are not met, we use all stars in the run to determine the offset
anyway, but flag the offset as unreliable.  675 of 764 runs (88\%) satisfy these criteria.

A rejection algorithm is applied to remove variable objects such as
variable stars and quasars that would degrade the calibration
measurements.  For each band, we reject objects with $\sigma > 0.05$
mag in that band, and require at least one good measurement in at
least 4 PS1 bands.  For the u-band comparison we also require $S/N>10$
in u-band.  After these cuts, the catalogs contain 24.7M, 47.5M,
60.0M, 61.8M, 59.3M stars, respectively, in the \emph{ugriz} fits.

\begin{table}
\begin{center}
\begin{tabular}{|r|rrrr|}
\hline
Band & $a_0$ & $a_1$ & $a_2$ & $a_3$ \\
\hline
$ u $ &  0.04438 & -2.26095 & -0.13387 &  0.27099 \\
$ g $ & -0.01808 & -0.13595 &  0.01941 & -0.00183 \\
$ r $ & -0.01836 & -0.03577 &  0.02612 & -0.00558 \\ 
$ i $ &  0.01170 & -0.00400 &  0.00066 & -0.00058 \\ 
$ z $ & -0.01062 &  0.07529 & -0.03592 &  0.00890 \\ 
$ y $ &  0.08924 & -0.20878 &  0.10360 & -0.02441 \\ 

\hline
\end{tabular}
\end{center}
\caption{Color transformation coefficients for the \texttt{qx\_noref} ubercalibration of the PV1 processing used in this work.}
\label{tbl:colortrans}
\end{table}

% ====================== Figure =======================================
  
\begin{figure*}
\begin{center}
\includegraphics[width=0.95\textwidth]{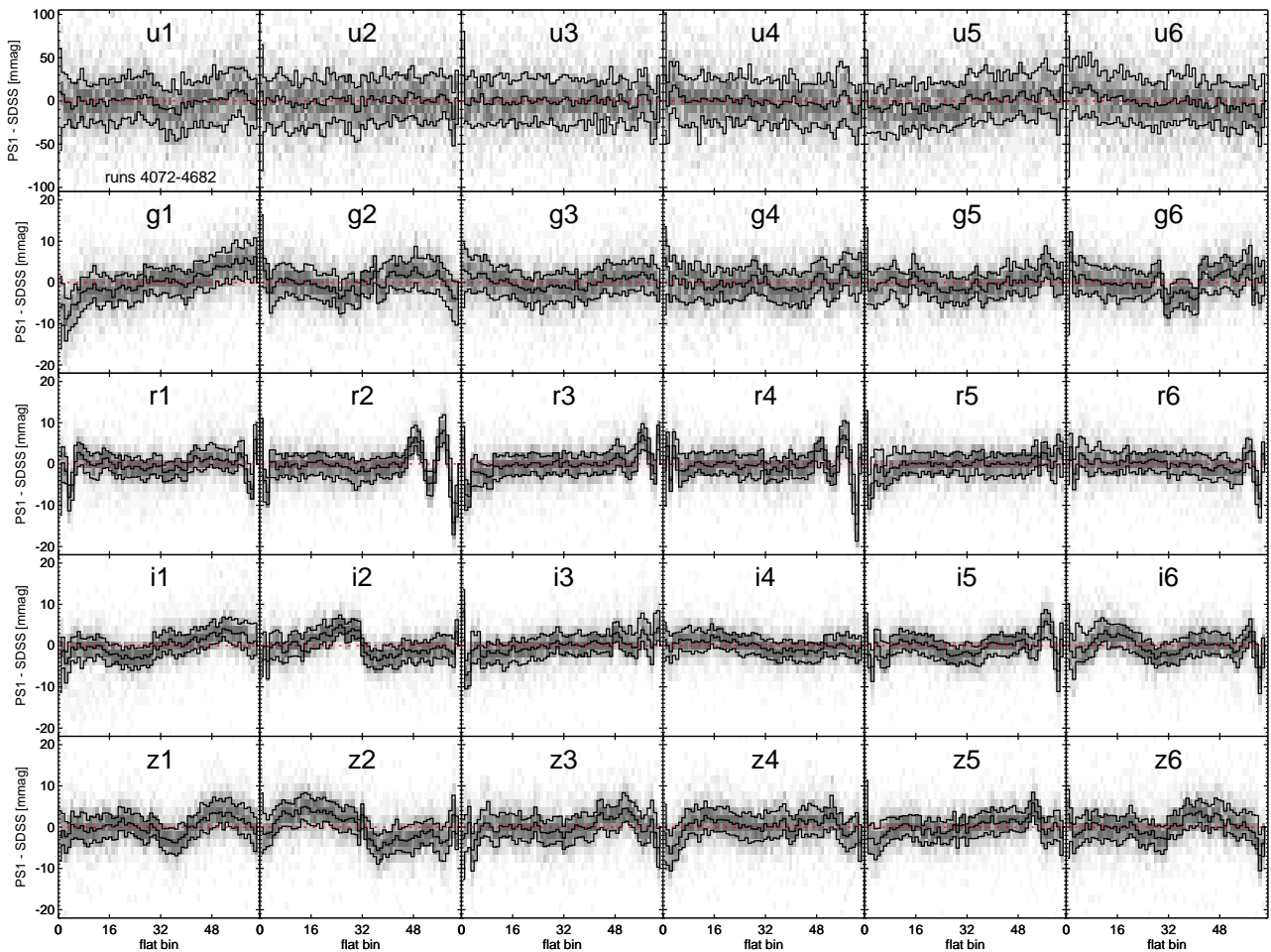}
\end{center}
\caption{Distribution of per-run flat residuals (PS1 minus SDSS DR8/DR9
  magnitudes) for the 2003-2004 observing season (runs 4072-4682) for each
  camera chip.  The flat residuals are measured in 64 bins of 32 CCD pixels
  each, and represented as a grayscale image with (16, 50, 84) percentile
  lines.  The oscillations in e.g. r2 and r4 are typical for these chips, and
  worsen as the
  survey progresses.  Padmanabhan et al. (2008) approximated the flat with a
  17-node B-spline, which could ring at this spatial scale.  The 5 mmag dip in
  g6 is constant until it disappears suddenly in 2005.  Jumps in the middle
  of two-amp chips often occur at the few mmag level (e.g. i2, z6).  The griz
  flats are stable to 2--3 mmag during a season, with u-band flats less well
  measured due to the extrapolation from \gps.  See supplemental
  materials for mean flats in every season.}
\label{fig:flat30}
\end{figure*}

\subsection{Color Transformations}
\label{sec:colortrans}
The SDSS $griz$ filters are similar to the \gps\rps\ips\zps\ filters,
but not identical.  We apply a correction to transform the PS1
magnitudes to the SDSS system, and then compare with SDSS magnitudes.
This transformation is defined as a function of PS1 magnitudes so that
it is stable as we alter the calibration of the SDSS magnitudes in
each run.  The transformation itself was determined from measurements
in PS1 Medium Deep Fields 9 and 10, which overlap SDSS stripe 82.  The
high redundancy in these fields in both surveys provides a very low
noise measurement of the color transformation parameters, with an RMS
per-star residual of 7 mmag.  We find that the color transformation is stable from year to year and field to field at the 3 mmag level. 
These transformations are a function of $g-i$ color, which behaves better
than transformations based on $g-r$ or $r-i$, as long as both
\gps\ and \ips\ are well measured.  For fainter stars, $g-r$ is better
for blue stars and $r-i$ is better for red stars, which may have no
$g$-band detection.  

The transformations are third-order polynomials in $x\equiv \gps-\ips$, with coefficients given in Table \ref{tbl:colortrans}.  
\begin{equation}
m_{p1} - m_{sdss} = a_0+a_1 x+ a_2 x^2 + a_3 x^3.
\end{equation}

They are valid for main-sequence stars with $0.4 < x < 2.7$.  Coefficients are provided for \gps$-u_{sdss}$ and
\yps$-z_{sdss}$ for completeness, with the caveat that these
extrapolations are much less reliable than the $griz$ transformations.
In particular, the extrapolation from PS1 colors to $u$ band is
strongly metallicity dependent, and should be used with caution.
The corrections are typically 0.01 mag in $r$ and $i$, up to 0.1 mag in $z$, and up to $0.25$ in $g$. 
These transformations, along with transformations as a function of
other colors and their inverses are presented by \cite{standards:2015}.

% ====================== Figure =======================================

\begin{figure*}[ht]
\begin{center}
\includegraphics[width=0.8\textwidth]{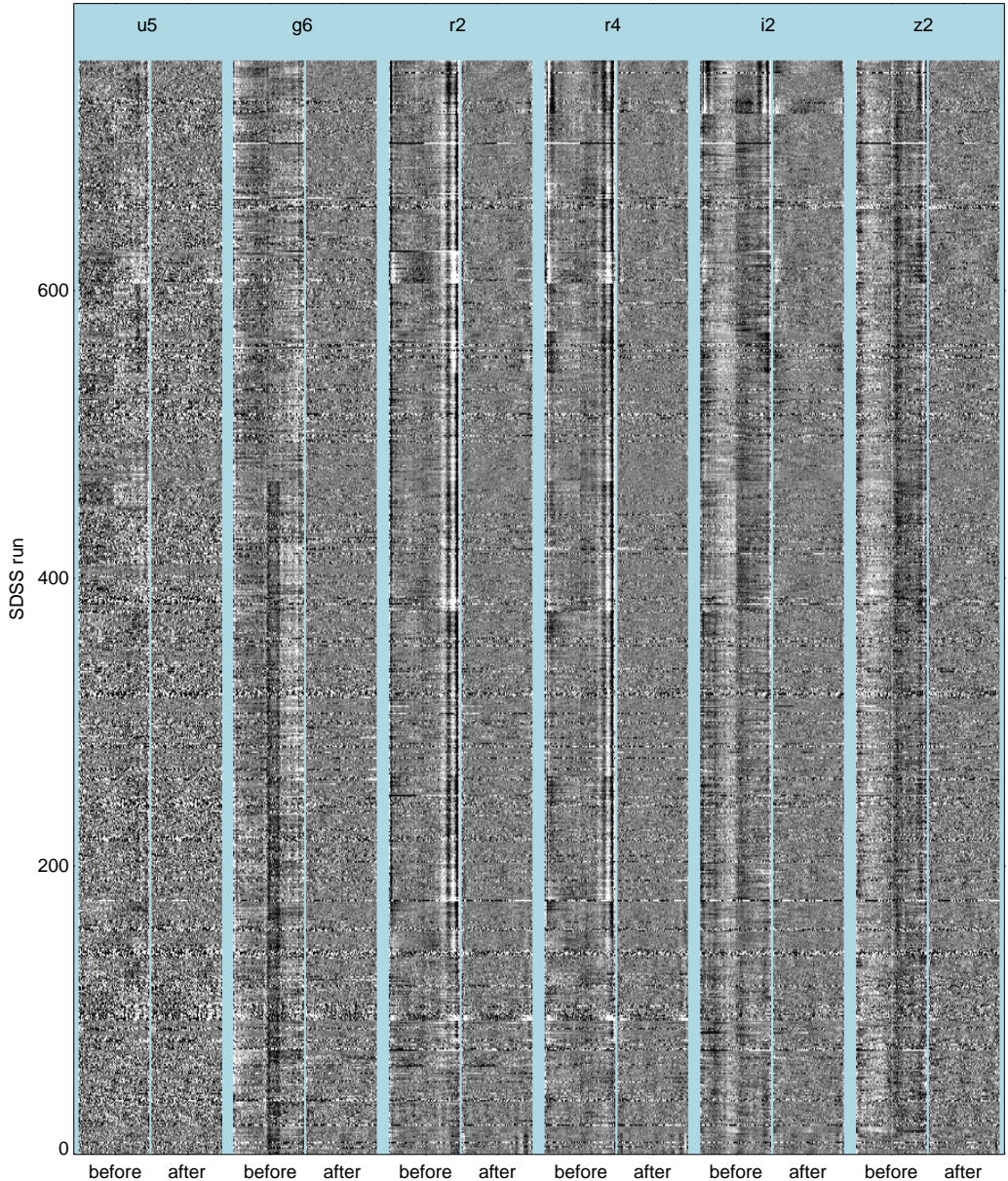}
\end{center}
\caption{SDSS flat residuals for 764 runs, represented as a grayscale from
  $-10$ to $+10$ mmag ($\pm50$ mmag for u-band).  Six of the 30 CCDs are
  shown.  Each row of pixels corresponds to the median flat of an SDSS run,
  and each pixel column corresponds to 32 camera pixel columns, ie. the 1-D
  flat for a 2048 pixel wide chip is represented by 64 bins.  In each case,
  the flat residuals ``before'' and ``after'' correction with the flat model of
  \S \ref{sec:flats} are shown.}
\label{fig:flat_ba}
\end{figure*}

% ====================== Figure =======================================
  
\begin{figure*}[ht]
\begin{center}
\includegraphics[width=0.9\textwidth]{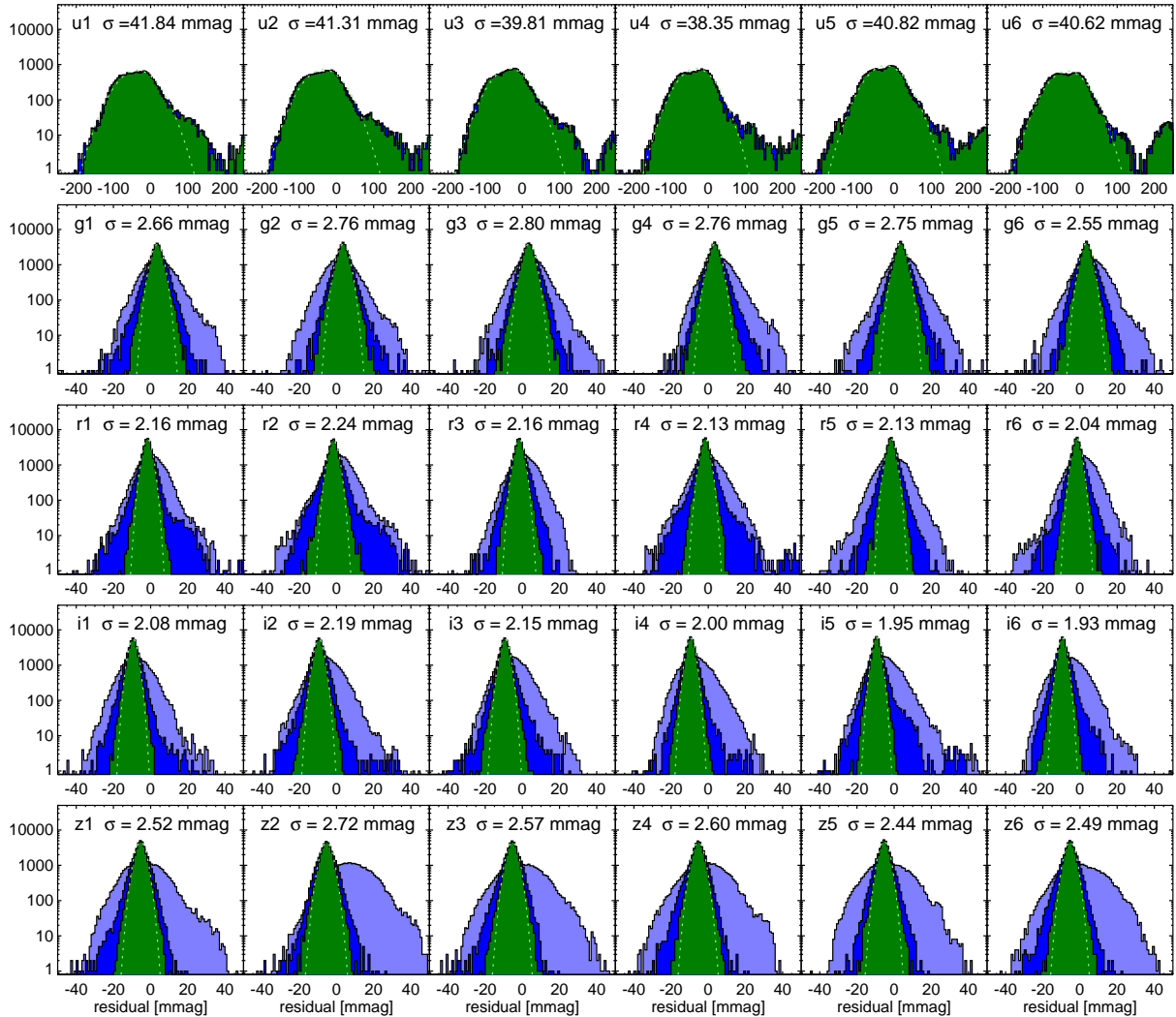}
\end{center}
\caption{Flat residual histograms:
the PS1-SDSS difference (\emph{light blue}), 
difference after subtracting mean difference per run (\emph{dark blue}),
and difference after applying flat corrections (\emph{green}), with
outlier-rejected Gaussian fit (\emph{dotted line}) with the RMS given.  The
flat correction does little to tighten the core of the distribution, but
dramatically suppresses outliers at $5\sigma$. 
}
\label{fig:flathist30}
\end{figure*}

% ====================== Figure =======================================

\begin{figure*}
\begin{center}
\includegraphics[width=0.8\textwidth]{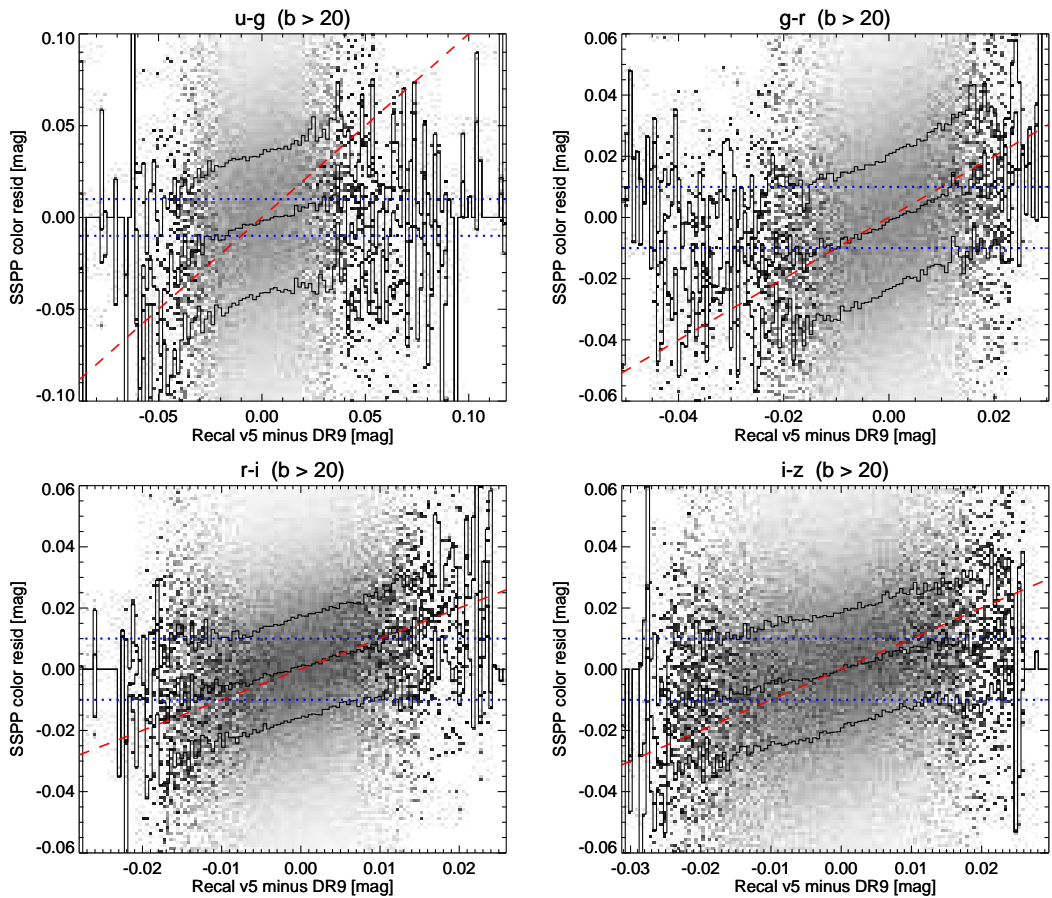}
\caption{SSPP calibration test using stars from \cite{Schlafly:2011}.  SSPP
  color residual is the color determined from spectral lines, minus DR8/9
  observed color, minus SFD dust reddening using SF11 coefficients.  Grayscale
  shows the distribution of this residual in small bins of recalibrated SDSS
  magnitude minus DR8/9, while black lines show the (16,50,84)th percentile. The
  dashed line has slope unity and y-intercept zero -- it is \emph{not} a fit.
  Stars with $b>20\degree$ are in good agreement except u-g, which is confused
  by metallicity gradients.  }
\label{fig:calibtest}
\end{center}
\end{figure*}
 
% ====================== Figure =======================================

\begin{figure*}
\begin{center}
\includegraphics[width=0.49\textwidth]{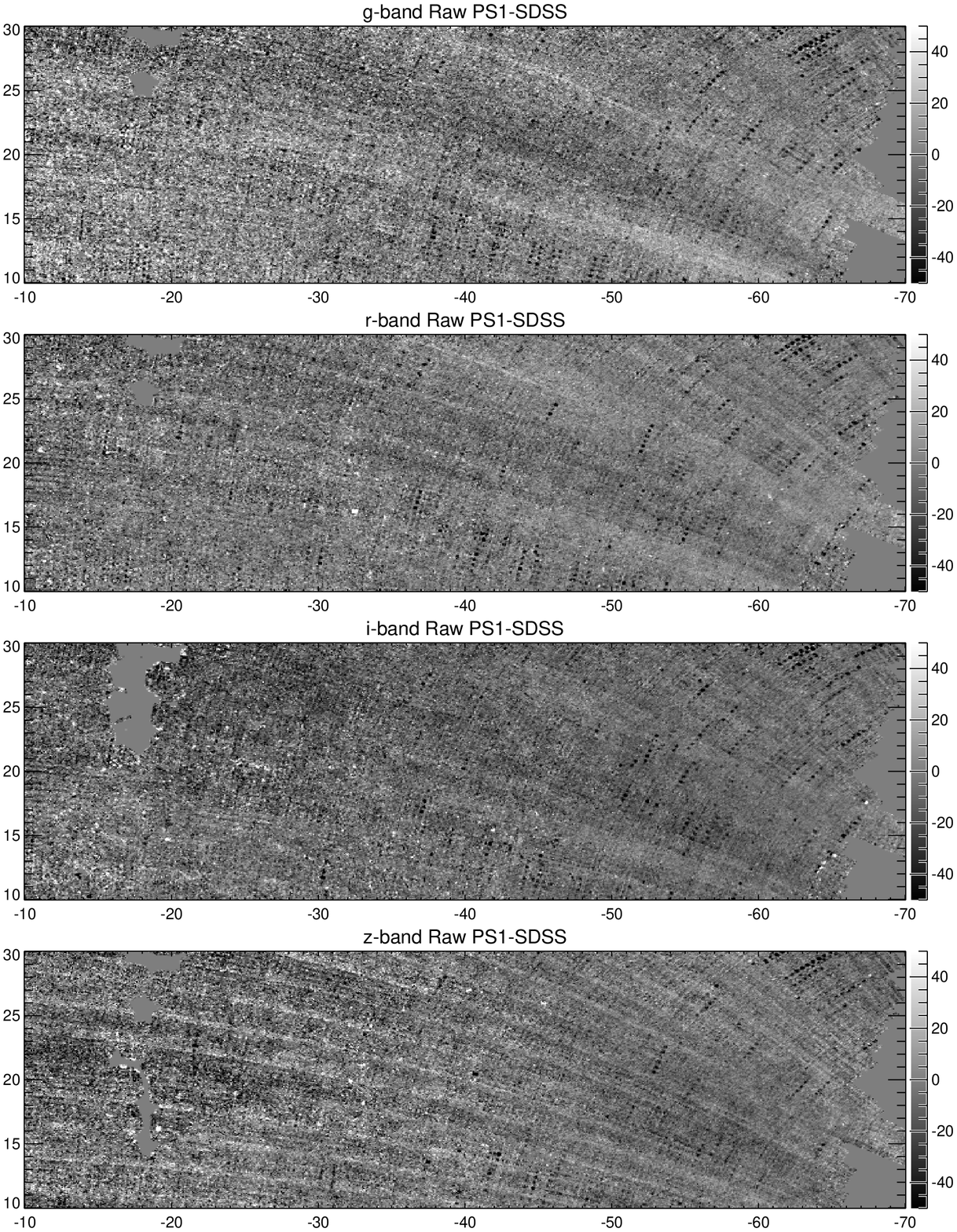}
\includegraphics[width=0.49\textwidth]{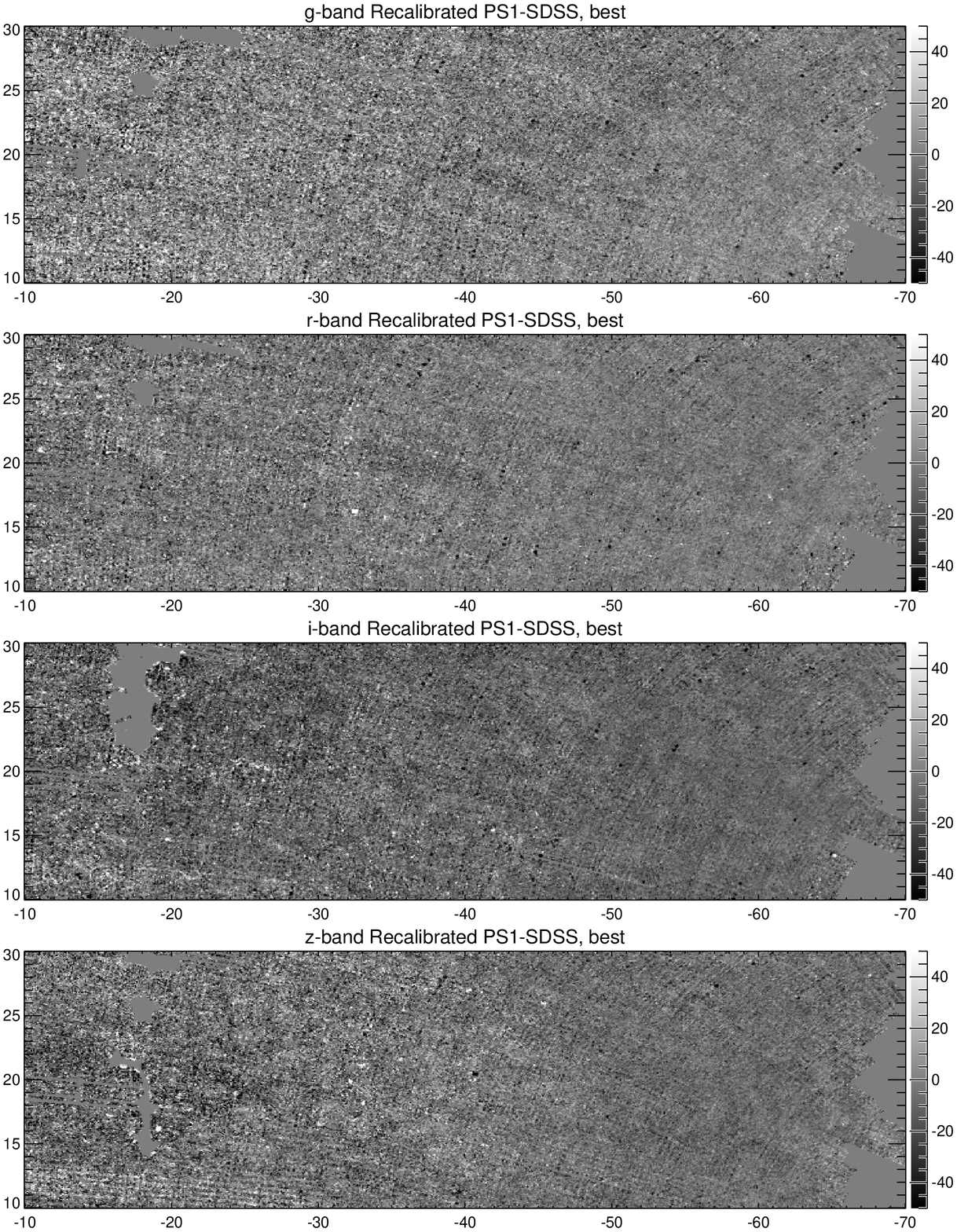}
\caption{Zoom on a patch of sky before (\emph{left}) and after (\emph{right})
  calibration and contrail rejection.  Stripes in the SDSS scan direction are
  mitigated by the recalibration.  Dark patches oriented in the cross-scan
  direction are contrails (see \S \ref{sec:contrails}).  White spots are
  usually poor photometry in single PS1 chips (see \S \ref{sec:ps1outliers}).  The $\sim 5$ mmag
  flatfield instability in PS1 manifests itself as a faint honeycomb pattern
  corresponding the pointings of the $3\degree$ diameter field of view.}
\label{fig:region1raw}
\end{center}
\end{figure*}

\section{A New Calibration}
\label{sec:calib}
\subsection{Run zero points}
\label{sec:runzp}

Using matched PS1 and SDSS detections of stars described in \refsec{match}, and
applying the color corrections described in \refsec{colortrans}, we
compute the PS1 minus SDSS difference for each SDSS detection.  A
grayscale map of this difference exhibits obvious stripes along SDSS
drift scans (Figure \ref{fig:healdiff5}).

For each SDSS band, we obtain the zero-point offset for each run by
computing the median difference.  This provides a robust estimate if
the atmospheric extinction is stable during the run.  The airmass
terms ($k$-terms) in the SDSS ubercal are generally well constrained,
but on nights with a small range in airmass they may be nearly
degenerate with that night's zero point.  In these cases, poorly
constrained $k$-terms can propagate to other areas of the SDSS
footprint that have low redundancy, such as survey edges or disjoint
regions.  Such problems affect a small fraction of the data (Figure \ref{fig:healdiff5}).

The SDSS ubercal was not able to fit a time variation in the airmass
term ($\dot{k})$ on a per-run basis.  However, the Photometric
Telescope \citep{hoggpt:2001} adjacent to the survey telescope
observed standard star fields continuously and obtained $k$ and
$\dot{k}$ for every night.  A systematic tendency for the atmosphere
to become more transparent during the night was found, and a mean
PT-derived $\dot{k}$ per band was included in the ubercal solution.
The $\dot{k}$ values of approximately 1 mmag/hr/airmass from the
earlier calibration \citep[Table 3 of][]{ubercal:2008} have been
preserved.

In order to minimize discrepancy with previous SDSS research, we have
adjusted the zero points derived above to preserve the median SDSS
calibration in the Galactic North.  In other words, if the color
transformations in Table \ref{tbl:colortrans} are used, the median $b
> 20\degree$ PS1 minus recalibrated SDSS difference is zero in $griz$.
The difference between north and south offsets is a few mmag in $g$
and increases in redder bands, reaching $13$ mmag in $z$ band (Table
\ref{tbl:northsouth}).  Assuming the PS1 offsets are correct, this
implies that SDSS magnitudes were too small (objects were too bright)
in the Galactic south relative to the north.

The $u$-band shift is strongly metallicity dependent and is therefore
not necessarily indicative of a photometric offset.  It has \emph{not}
been applied and is uncertain by several hundredths of a magnitude. 

\begin{table}
\begin{center}
\begin{tabular}{|c|r|r|r||}
\hline
band & North & South & North-South \\
     & (mmag) & (mmag) & (mmag)  \\
\hline
u  & $ -45.10 $ & $ -18.81 $ & $ -26.29 $ \\
g  & $   1.18 $ & $   3.45 $ & $  -2.27 $ \\
r  & $  -2.14 $ & $   2.71 $ & $  -4.85 $ \\
i  & $  -9.15 $ & $  -1.29 $ & $  -7.86 $ \\
z  & $  -5.24 $ & $   7.41 $ & $ -12.66 $ \\
\hline
\end{tabular}
\end{center}
\caption{Median PS1 minus SDSS magnitudes in the North ($b >
  20\degree$), South ($b < -20\degree$), and the difference, based on
  the color transformations given in \refsec{colortrans}.  These
  shifts are with respect to the median in the MD09 and MD10 reference
  fields, where the color transformations were originally determined.
  The $griz$ North offsets are included in the color terms in Table
  \ref{tbl:colortrans} (see \S \ref{sec:runzp}).}
\label{tbl:northsouth}
\end{table}

\subsection{Flat fields}
\label{sec:flats}
The updated zero points reduce the stripe residuals along SDSS scans
(Figure \ref{fig:healdiff5}), but significant
smaller structure remains, motivating an examination of the SDSS
flats.  The SDSS drift scan technique averages over pixel rows on each CCD, making
the flat a 1-D function of pixel column.  We understand the flat to
include pixel sensitivity, filter response, and amplifier gains.  Most
SDSS chips have 2 amplifier readout, and they are usually stable to
within $\sim 1$ mmag but in some cases are observed to jump suddenly
by $\sim 5$ mmag with respect to each other.

For each run+chip, the differences are sorted into 64 bins in CCD pixel
column (32 pixels per bin) to determine the \emph{flat residual}, $F_r$.
$F_r$ is simply the median difference for ``good'' stars in each of the 64
bins.  We distinguish between the observed flat residual and the \emph{flat
  correction}, which is a model of it to be applied in the recalibration.

The flat residuals are generally stable in time, but with some sudden
jumps.  These jumps usually occur on the ``season'' boundaries
established previously \citep{ubercal:2008}, although we use a subset
of these seasons: 8 seasons ending at MJD 51251, 51865, 52144, 52872,
53243, 53959, 55090, and 55153.  These boundaries correspond to run
numbers 725, 1869, 2504, 4069, 4792, 6245, 8032, and 8162. We compute
a median flat correction $F_s$ per season/chip, and use that as a
basis for the flat model.  In each season, the residuals $F_r$
scatter about $F_s$ with an RMS of 2-3 mmag (Figure \ref{fig:flat30}).

This does not remove all the structure in the flat residuals, so we
model the remaining structure with a principal component expansion,
one chip at a time (see Figure \ref{fig:flat_ba} for a comparison of
the residuals before and after correction).  We compute
principal components of the residuals (after subtracting the season
medians) for all runs with more than 10 good stars (according to
CALIB\_STATUS bit 0) in each of the 64 camcol bins, a criterion that
rejects runs without significant photometric data.  For each run, we fit coefficients for
the first 4 principal components and add these components to the
per-season flat correction for that run.  For short or unphotometric runs that do not
have 10 good stars in each 32 pixel camcol bin, we simply apply the
per-season flat correction and record this choice in CALIB\_STATUS bit
13.  If the RMS residual for a run is less than twice the median RMS
residual for all runs, the run is deemed to have low noise, and we set
CALIB\_STATUS bit 14 (see Table \ref{tbl:calib_status}).  We repeated
this entire procedure using 1-6 PC coefficients, and found by inspection that
4 PCs were adequate to remove the apparent structure in the flat
residuals.

These models yield residuals 
$\sigma_{rej}$ of $2-3$ mmag in $griz$ bands (Figure \ref{fig:flathist30}).  The
distribution has many-sigma outliers at the 1\% level, but they are in all
cases $< 20$ mmag.  Subject to the assumption that the flat is constant during
a run, this implies that the flat is known at the $\sim 3$ mmag level in
\emph{griz} and 15 mmag in \emph{u}.

The extrapolation from $\gps$ to $u_{sdss}$ is poorly defined, because
of its dependence on stellar metallicity and dust extinction.  Because
metallicity gradients are on much larger scales than the features in
the flat residuals and dust is uncorrelated with them, it is expected
that we can get a good $u$-band flat anyway.  We apply the updated
$u$-band flat corrections, but do not alter the $u$-band zero points.

\begin{figure*}
\begin{center}
\includegraphics[width=0.45\textwidth]{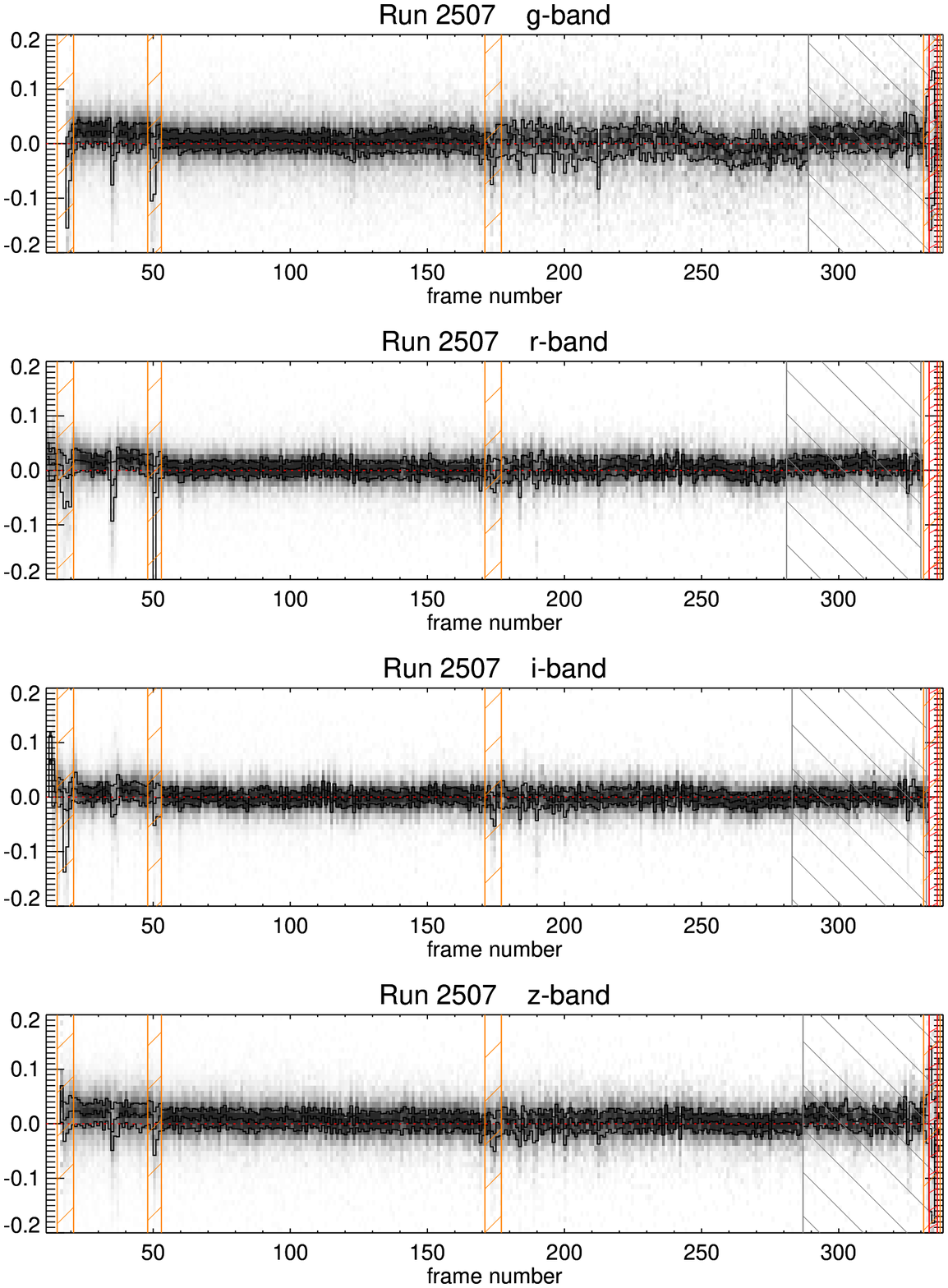}
\includegraphics[width=0.45\textwidth]{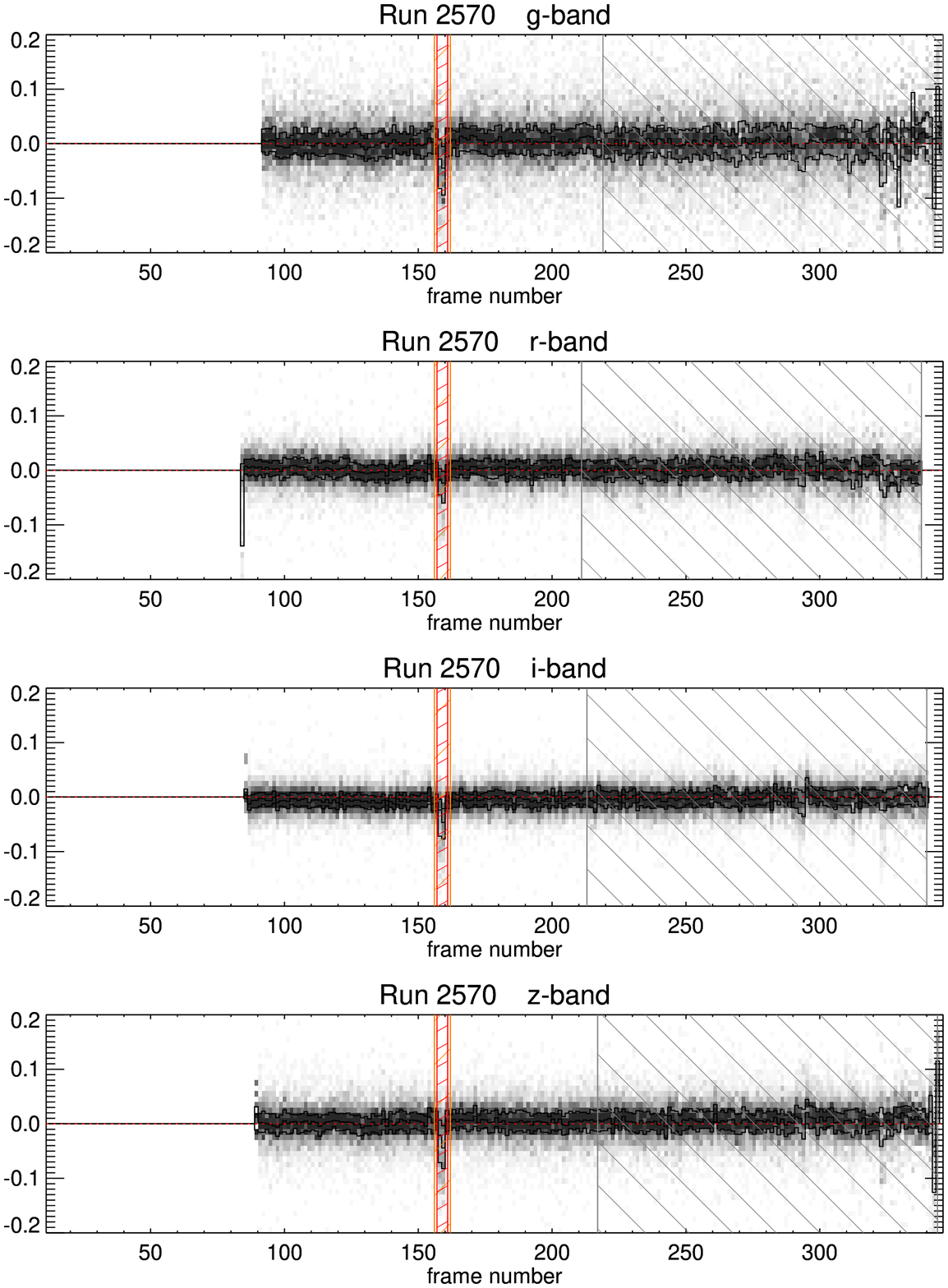}
\vskip 0.25in
\includegraphics[width=0.45\textwidth]{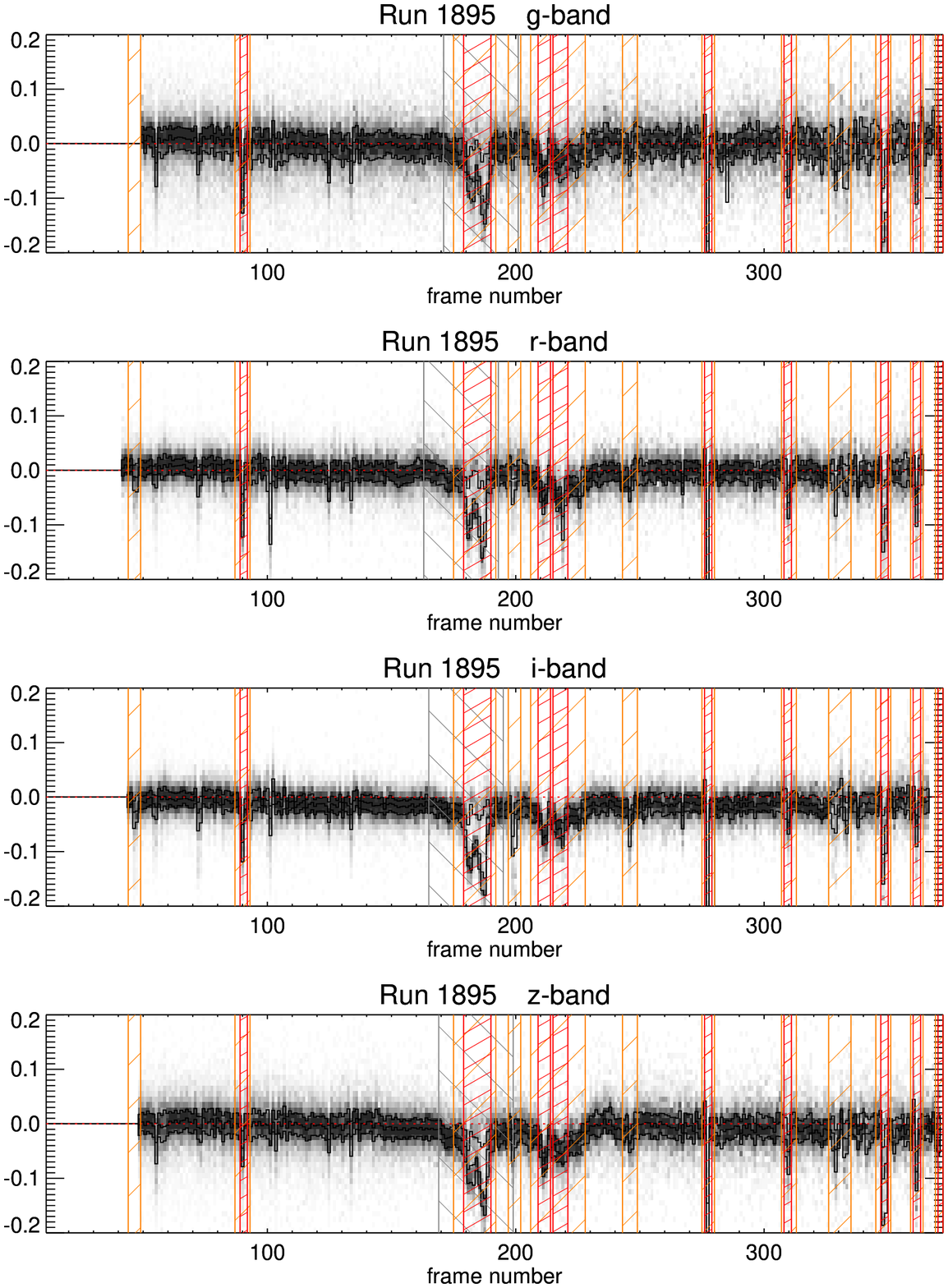}
\includegraphics[width=0.45\textwidth]{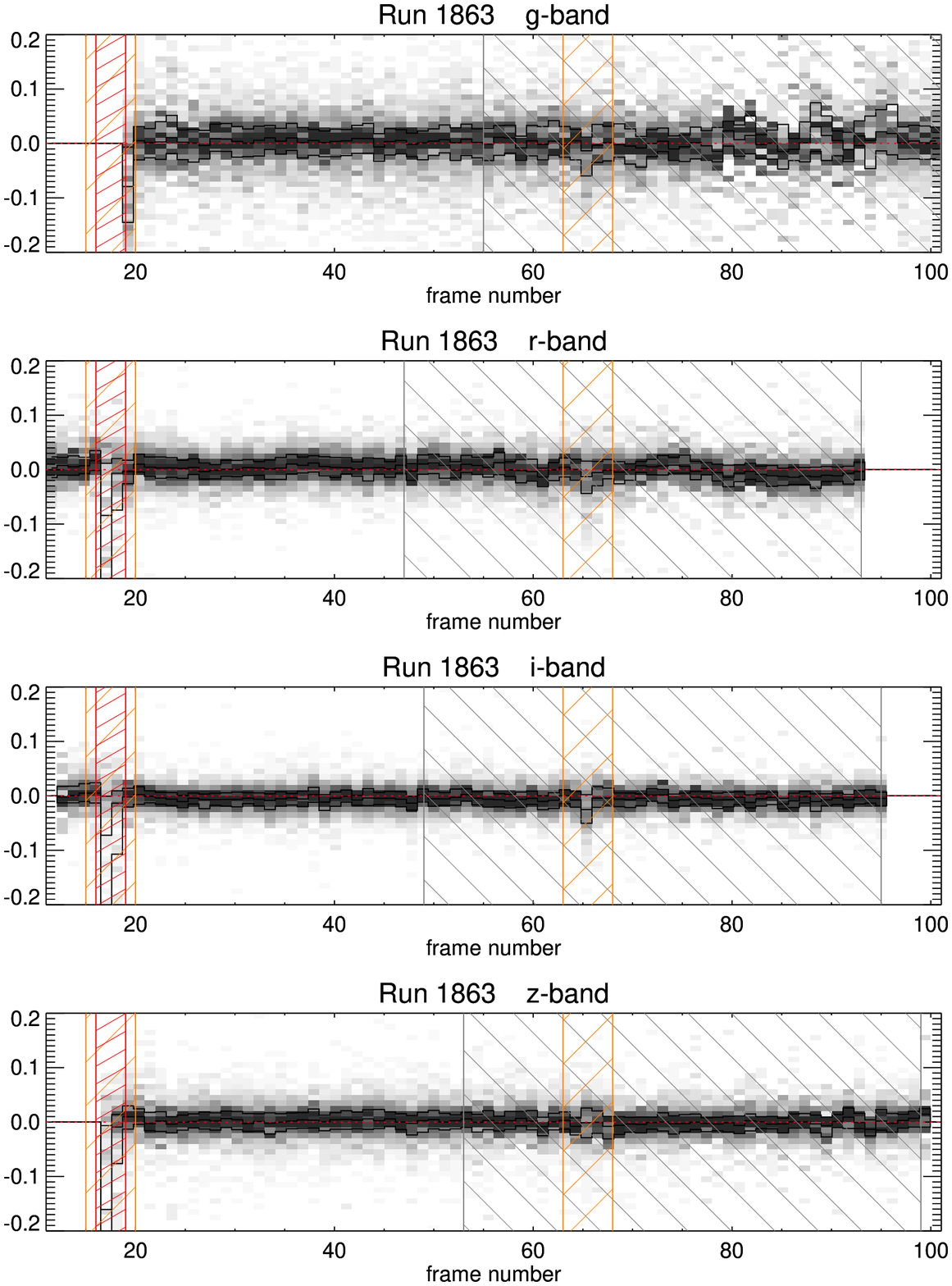}
\caption{PS1 minus SDSS magnitudes as a function of SDSS frame number for 4
  SDSS runs, as described in \S \ref{sec:contrails}.  We flag deviations
  aligned in SDSS observation time as contrails, padding 2 frames before and
  after.  Periods that are unphotometric (red hatch) or 
  ``sightly unphotometric'' (orange hatch) are indicated.  
  Most runs exhibit excellent agreement with PS1, showing
  only occasional deviations (upper left panels).  In some cases, a deviation
  is present in the first or last frame (upper right panels).  In other cases,
  a run may show many deviations (lower left), only some of which are marked
  bad in the SDSS \texttt{CALIB\_STATUS} flags (gray hatch).  In run 1863 (lower
  right), unphotometric data were tied to overlapping runs starting at
  field 47, with good results.  }
\label{fig:contrails}
\end{center}
\end{figure*}

\subsection{Validation}
As an independent validation of our calibration, we use a
spectroscopic sample of stars for which we have reddening estimates.
The SDSS took spectra of over 1 million objects, including over
250,000 stars.  The SDSS SEGUE Stellar Parameter Pipeline
\citep[SSPP;][]{Lee:2008} returns
several estimates of the stellar type, including one based on the
continuum-normalized spectrum, effectively using only line information
and no information from the continuum or the \emph{ugriz} photometry.
This estimate is not affected by dust reddening or
photometric calibration errors.  \cite{Schlafly:2011} used these stellar types for
261,496 stars, along with appropriate spectral libraries, to estimate
the true color of each star and subtract it from each broadband color
to obtain a color excess.  They interpreted the color excesses as dust
reddening, and used them to argue that the \cite{Fitzpatrick:1999}
reddening law is a good fit to the data, and to derive new
calibrations for the \cite{SFD:1998} map in 88 bands \citep[][Table 6]{Schlafly:2011}.

We use this same stellar sample to validate our current work.  SDSS photometric
calibration residuals also appear in the SSPP color residual, and we expect
those residuals to be positively correlated with the correction we derive in
\S \ref{sec:calib}.  Because the per-star scatter is large, it is useful to bin in
our correction (recalibrated magnitude minus DR8/DR9 magnitude) and plot the
median and 16th,84th percentiles (Figure \ref{fig:calibtest}).  The median
line in each panel has a slope of $\sim1/2$ in u-g and close to 1 in the other
colors.  Any noise in recalibrated color will spread the data points
horizontally and cause the slope to be less than 1.  It is not surprising that
u-g is worse in this regard, since we did not apply zero-point corrections to
u band.  In the other bands, the PS1-based corrections are highly correlated
with the correction implied by the SSPP sample.

\section{Contrails}
\label{sec:contrails}
PS1 and SDSS attempt to censor unphotometric data, both during
observations and later during calibration.  Inevitably, some periods of
non-photometricity persist in both data sets.  Condensation trails from
airplanes (``contrails'') are especially pernicious, because they usually
compromise photometry for only a single exposure (in PS1) or a frame or two of
drift scan (in SDSS).  In the following, we identify brief periods of
unphotometric data in SDSS, and without loss of generality refer to them as
contrails.

\subsection{SDSS outliers}
In Figure \ref{fig:region1raw}, dark patches are apparent, representing
regions where SDSS stars are too faint relative to PS1.  In most cases, these
appear in groups of 5 or 6 in the cross-scan direction, with the same
quarter-degree spacing as the SDSS camera columns.  It is also clear from
Figure \ref{fig:region1raw} that similar features appear in multiple bands
with slight offsets, consistent with the spacing between the \emph{griz} chips
in the SDSS camera.  For example, $i$ is observed 72 seconds after $r$, so a
simultaneous deviation appears separated by $18'$ on the sky.  The bands are
always observed in \emph{riuzg} order. 

For all 764 science runs in DR8/DR9, we compute the PS1 minus SDSS
difference (including color terms) as a function of SDSS frame number
for $griz$ bands.\footnote{SDSS runs are split into \emph{frames}
  according to observation time, and then offsets are applied to
  obtain \emph{field} numbers corresponding to locations on the sky.
  In DR8/DR9, most runs begin with field 11, which is frame
  (15,19,11,13,17) in (u,g,r,i,z) bands.  Field 11 is generally the
  first field that has complete data in all five filters, and earlier
  fields (1--10) are discarded.}  We flag deviations aligned in SDSS
observation time as either ``unphotometric'' or ``slightly
unphotometric'', and set CABLIB\_STATUS bits 6 or 7,
respectively.  When all 4 bands are present ($griz$), we label a frame
unphotometric if the median deviation in that frame is greater than
0.04 mag in at least two bands.  If only 3 bands are present (e.g. near
the beginning or end of a run), a single-band deviation of 0.04 mag is
sufficient to be unphotometric.  In either case, we pad the masked region
be 1 frame.  We recommend avoiding the masked frames
for any work requiring accurate photometry. 

For users desiring a more pristine sample, we define the ``slightly
unphotometric'' mask bit.  This bit is set when a frame has a
deviation of at least 0.015 mag in 3 bands, or 0.025 in at least 2
bands.  These criteria were chosen
by inspection, and strike a balance between catching most of the
outliers and avoiding most false positives.  As above, when only 3
bands are available, the criteria are loosened by one band.  This
loosening increases the chance of a false-positive at the beginning or
end of a run, which is sensible given that a run might end because
clouds rolled in.  In either case, the ``slightly unphotometric''
region is then padded by 2 frames.  

Figure \ref{fig:contrails} displays the PS1-SDSS residual vs. time for four of
the 764 runs.  Most runs exhibit excellent agreement with PS1, with at most
one or two contrails, and often none.  In some cases (e.g. run 1895) several
unphotometric periods do not coincide with the SDSS
\texttt{CALIB\_STATUS} photometricity flag.  In other cases, (e.g. run 1863), SDSS has
been too cautious, and has marked data unphotometric that appears to be fine
in our comparison.

\subsection{PS1 outliers}
\label{sec:ps1outliers}
The difference maps also contain outliers in the opposite sense, but
these light splotches are the size of a PS1 chip, suggesting PS1 stars
on a chip are too faint.  In previous iterations of this analysis,
most such stars were on chips 14, 66, and to a lesser extent 27.
Internal PS1 comparisons find these chips to be less reliable than the
others.

In the latest iteration of PS1 ubercal (version \texttt{qx\_noref}) we include
some outlier rejection to discard an exposure of a chip if it
disagreed with consensus by more than some threshold.  This seems to
have correctly rejected chips 27 and 66 in most cases, at least where we
have enough coverage.  However chip 14 still leaks through, because it is
partly masked, and in some cases half of the chip is good, so it does
not trigger the outlier rejection.  This will be addressed in the
final PS1 calibration, but is of no consequence for this work.

%\section{An Application : Large Scale Structure} % (fold)
%\label{sec:app}

% section An Application : Large Scale Structure (end)

\section{Conclusions} % (fold)
\label{sec:discuss}
The SDSS photometry has been used in thousands of papers and is one
of the most valuable astronomical data sets.  It is the basis 
of target selection for all of the SDSS spectroscopy (including the 
e-BOSS survey), and it is a well studied data set for
extended source photometry and colors.  For many applications, 
precise calibration is increasingly important, and the recalibrated 
SDSS is more stable than either PS1 or SDSS alone. 

This work presents a new calibration, based on the first 3 years of
Pan-STARRS1 $3\pi$ photometry.  PS1 observes the sky with much higher
redundancy than SDSS and frequent observations of the Medium Deep
Fields add rigidity to the photometric solution.  By comparing over 60
million SDSS detections with PS1, we derive new zero points for each
$griz$ chip in each SDSS run, and determine flat-field corrections at
the 3 mmag level in $griz$ and 15 mmag in $u$.  We assess the
stability of the new correction on the scale of an SDSS field 
($ \sim 13.5'$) by binning the SDSS minus PS1 difference in HEALPix 
$ N_{\rm side}=256$ pixels ($\sim 13.7'$).  Using these new calibration
parameters (and appropriate bandpass corrections) the difference has
an RMS of \{9,7,7,8\} mmag in $griz$.  In the limit of high stellar
density ($> 300$ stars per $15'$ pixel) the RMS asymptotes to
\{7.5,6.3,6.1,7.2\} mmag in $griz$.  However, these regions also have
more overlap between SDSS runs, so we take \{9,7,7,8\} mmag as
representative.  On much smaller scales $N_{\rm side}=1024 ~(3.4')$,
the RMS asymptotes to \{16,12,12,14\} mmag, but this likely includes a
significant contribution from residuals in the PS1 focal plane.

In principle, a cross calibration of SDSS and PS1 could be performed, 
solving for the calibration parameters of both surveys simultaneously.
We have resisted this temptation for two reasons:

The PS1 calibration, with a single atmospheric $k$-term and zero
point per band per night, is already formally tightly constrained and including
constraints from SDSS would add little.  The PS1 model could be 
generalized to include more freedom and many more parameters.  However,
the PS1 zero points and flats are already so good that a substantial 
part of the photometric error in psf flux estimates comes from errors
in the psf at each location in each exposure.  We are reluctant to 
treat these errors as pure photometric offsets, as they depend on 
object size and shape. 

Even with an expanded PS1 calibration model with enough freedom that
SDSS helps constrain it, we would have a qualitatively different
photometric stability inside the SDSS footprint compared to the rest
of the sky.  It is more appealing to have PS1 be a monolithic survey
with uniform properties across 3/4 of the sky.  However, we anticipate
a more general approach in future surveys, in which a simultaneous
solution for calibration parameters of multiple large data sets might
be computed.  Such an approach is optimal, \emph{if} a calibration
model can be formulated that jointly describes the data sets to the
required level of detail.

The new flats, per-run zero points, and a mask of short periods of
non-photometricity (e.g contrails) are encoded in the
\texttt{calibPhotomGlobal} files and are publicly
available.\footnote{\url{http://faun.rc.fas.harvard.edu/ps1sdss/dr9/calib/v5b}}
They are expected to propagate into SDSS data release 13 (summer, 2016), but may
be used immediately via the procedure \texttt{sdss\_recalibrate},
available in the SDSS3 \texttt{IDLUTILS} repository.\footnote{\url{http://www.sdss3.org/dr8/software/idlutils.php}}

% section discuss (end)

\vskip 0.15in {\bf \noindent Acknowledgments:} 

We acknowledge helpful conversations with Michael
Blanton. D.P.F. and E.F.S. have been partially supported by NASA grant
NNX10AD69G.  E.F.S acknowledges funding by Sonderforschungsbereich SFB
881 ``The Milky Way System'' (subproject A3) of the German Research
Foundation (DFG).
This research made use of the NASA Astrophysics Data System (ADS) and the IDL
Astronomy User's Library at Goddard.\footnote{Available at
  \url{http://idlastro.gsfc.nasa.gov}}

% SDSS boilerplate
Funding for SDSS-III has been provided by the Alfred P. Sloan Foundation, the
Participating Institutions, the National Science Foundation, and the
U.S. Department of Energy Office of Science. The SDSS-III web site is
\texttt{http://www.sdss3.org/}.

SDSS-III is managed by the Astrophysical Research Consortium for the
Participating Institutions of the SDSS-III Collaboration including the
University of Arizona, the Brazilian Participation Group, Brookhaven
National Laboratory, Carnegie Mellon University, University of Florida, the
French Participation Group, the German Participation Group, Harvard
University, the Instituto de Astrofisica de Canarias, the Michigan State/Notre
Dame/JINA Participation Group, Johns Hopkins University, Lawrence Berkeley
National Laboratory, Max Planck Institute for Astrophysics, Max Planck
Institute for Extraterrestrial Physics, New Mexico State University, New York
University, Ohio State University, Pennsylvania State University, University
of Portsmouth, Princeton University, the Spanish Participation Group,
University of Tokyo, University of Utah, Vanderbilt University, University
of Virginia, University of Washington, and Yale University.

%PS1 boilerplate
The Pan-STARRS1 Surveys (PS1) have been made possible through contributions of
the Institute for Astronomy, the University of Hawaii, the Pan-STARRS
Project Office, the Max-Planck Society and its participating institutes, the
Max Planck Institute for Astronomy, Heidelberg and the Max Planck Institute
for Extraterrestrial Physics, Garching, The Johns Hopkins University,
Durham University, the University of Edinburgh, Queen’s University Belfast,
the Harvard-Smithsonian Center for Astrophysics, the Las Cumbres Observatory
Global Telescope Network Incorporated, the National Central University of
Taiwan, the Space Telescope Science Institute, the National Aeronautics and
Space Administration under Grant No. NNX08AR22G issued through the Planetary
Science Division of the NASA Science Mission Directorate, the National
Science Foundation under Grant No. AST-1238877, and the University of
Maryland.

\section{Supplemental material}
The QA plots generated by this study go far beyond the scope of this paper.
We provide supplemental plots,\footnote{\url{http://faun.rc.fas.harvard.edu/ps1sdss/plots/v5b}} 
including the following:

\begin{itemize}
\item \texttt{flat30-all}     As in Fig. \ref{fig:flat30}, but for all seasons. 
\item \texttt{sdss\_contrails}  - griz contrail plots for 764 runs
\item \texttt{healdiff}  Full-sky maps at healpix nside=256 (15 arcmin pixels)
\item \texttt{healdiff} and Nside=1024 (3.5 arcmin pixels)
\item \texttt{all\_pdfs.tar} containing all of the above and more. 
\end{itemize}

%\newpage
%
%\section{Todo list}
%
%\begin{itemize}
%\item What about aperture mags?
%\item Should we explicitly comment on color calibration?
%\item INTRO: Should we motivate the paper with some science applications?
%\item How do we flag an entire run as failed?
%\item What criteria make a magnitude ok in Sec 2?

%\end{itemize}

\newpage
\nocite{*}
\bibliographystyle{apj}
\bibliography{ps1sdss}

\begin{thebibliography}{32}
\expandafter\ifx\csname natexlab\endcsname\relax\def\natexlab#1{#1}\fi

\bibitem[{{Ahn} {et~al.}(2012){Ahn}, {Alexandroff}, {Allende Prieto},
  {Anderson}, {Anderton}, {Andrews}, {Aubourg}, {Bailey}, {Balbinot}, {Barnes},
  \& et~al.}]{DR9}
{Ahn}, C.~P., {et~al.} 2012, \apjs, 203, 21

\bibitem[{{Ahn} {et~al.}(2013){Ahn}, {Alexandroff}, {Allende Prieto}, {Anders},
  {Anderson}, {Anderton}, {Andrews}, {Aubourg}, {Bailey}, {Bastien}, \&
  et~al.}]{DR10}
---. 2013, ArXiv e-prints

\bibitem[{{Aihara} {et~al.}(2011){Aihara}, {Allende Prieto}, {An}, {Anderson},
  {Aubourg}, {Balbinot}, {Beers}, {Berlind}, {Bickerton}, {Bizyaev}, {Blanton},
  {Bochanski}, {Bolton}, {Bovy}, {Brandt}, {Brinkmann}, {Brown}, {Brownstein},
  {Busca}, {Campbell}, {Carr}, {Chen}, {Chiappini}, {Comparat}, {Connolly},
  {Cortes}, {Croft}, {Cuesta}, {da Costa}, {Davenport}, {Dawson}, {Dhital},
  {Ealet}, {Ebelke}, {Edmondson}, {Eisenstein}, {Escoffier}, {Esposito},
  {Evans}, {Fan}, {Femen{\'{\i}}a Castell{\'a}}, {Font-Ribera}, {Frinchaboy},
  {Ge}, {Gillespie}, {Gilmore}, {Gonz{\'a}lez Hern{\'a}ndez}, {Gott}, {Gould},
  {Grebel}, {Gunn}, {Hamilton}, {Harding}, {Harris}, {Hawley}, {Hearty}, {Ho},
  {Hogg}, {Holtzman}, {Honscheid}, {Inada}, {Ivans}, {Jiang}, {Johnson},
  {Jordan}, {Jordan}, {Kazin}, {Kirkby}, {Klaene}, {Knapp}, {Kneib},
  {Kochanek}, {Koesterke}, {Kollmeier}, {Kron}, {Lampeitl}, {Lang}, {Le Goff},
  {Lee}, {Lin}, {Long}, {Loomis}, {Lucatello}, {Lundgren}, {Lupton}, {Ma},
  {MacDonald}, {Mahadevan}, {Maia}, {Makler}, {Malanushenko}, {Malanushenko},
  {Mandelbaum}, {Maraston}, {Margala}, {Masters}, {McBride}, {McGehee},
  {McGreer}, {M{\'e}nard}, {Miralda-Escud{\'e}}, {Morrison}, {Mullally},
  {Muna}, {Munn}, {Murayama}, {Myers}, {Naugle}, {Neto}, {Nguyen}, {Nichol},
  {O'Connell}, {Ogando}, {Olmstead}, {Oravetz}, {Padmanabhan},
  {Palanque-Delabrouille}, {Pan}, {Pandey}, {P{\^a}ris}, {Percival},
  {Petitjean}, {Pfaffenberger}, {Pforr}, {Phleps}, {Pichon}, {Pieri}, {Prada},
  {Price-Whelan}, {Raddick}, {Ramos}, {Reyl{\'e}}, {Rich}, {Richards}, {Rix},
  {Robin}, {Rocha-Pinto}, {Rockosi}, {Roe}, {Rollinde}, {Ross}, {Ross},
  {Rossetto}, {S{\'a}nchez}, {Sayres}, {Schlegel}, {Schlesinger}, {Schmidt},
  {Schneider}, {Sheldon}, {Shu}, {Simmerer}, {Simmons}, {Sivarani}, {Snedden},
  {Sobeck}, {Steinmetz}, {Strauss}, {Szalay}, {Tanaka}, {Thakar}, {Thomas},
  {Tinker}, {Tofflemire}, {Tojeiro}, {Tremonti}, {Vandenberg}, {Vargas
  Maga{\~n}a}, {Verde}, {Vogt}, {Wake}, {Wang}, {Weaver}, {Weinberg}, {White},
  {White}, {Yanny}, {Yasuda}, {Yeche}, \& {Zehavi}}]{DR8}
{Aihara}, H., {et~al.} 2011, \apjs, 193, 29

\bibitem[{{Finkbeiner} {et~al.}(2016)}]{standards:2015}
{Finkbeiner}, D.~P., {et~al.} 2016, \apj, \emph{in prep}, 000

\bibitem[{{Fitzpatrick}(1999)}]{Fitzpatrick:1999}
{Fitzpatrick}, E.~L. 1999, \pasp, 111, 63

\bibitem[{{Fukugita} {et~al.}(1996){Fukugita}, {Ichikawa}, {Gunn}, {Doi},
  {Shimasaku}, \& {Schneider}}]{Fukugita:1996}
{Fukugita}, M., {Ichikawa}, T., {Gunn}, J.~E., {Doi}, M., {Shimasaku}, K., \&
  {Schneider}, D.~P. 1996, \aj, 111, 1748

\bibitem[{{Gunn} {et~al.}(1998){Gunn}, {Carr}, {Rockosi}, {Sekiguchi}, {Berry},
  {Elms}, {de Haas}, {Ivezi{\'c}}, {Knapp}, {Lupton}, {Pauls}, {Simcoe},
  {Hirsch}, {Sanford}, {Wang}, {York}, {Harris}, {Annis}, {Bartozek},
  {Boroski}, {Bakken}, {Haldeman}, {Kent}, {Holm}, {Holmgren}, {Petravick},
  {Prosapio}, {Rechenmacher}, {Doi}, {Fukugita}, {Shimasaku}, {Okada}, {Hull},
  {Siegmund}, {Mannery}, {Blouke}, {Heidtman}, {Schneider}, {Lucinio}, \&
  {Brinkman}}]{Gunn:1998}
{Gunn}, J.~E., {et~al.} 1998, \aj, 116, 3040

\bibitem[{{Gunn} {et~al.}(2006){Gunn}, {Siegmund}, {Mannery}, {Owen}, {Hull},
  {Leger}, {Carey}, {Knapp}, {York}, {Boroski}, {Kent}, {Lupton}, {Rockosi},
  {Evans}, {Waddell}, {Anderson}, {Annis}, {Barentine}, {Bartoszek}, {Bastian},
  {Bracker}, {Brewington}, {Briegel}, {Brinkmann}, {Brown}, {Carr},
  {Czarapata}, {Drennan}, {Dombeck}, {Federwitz}, {Gillespie}, {Gonzales},
  {Hansen}, {Harvanek}, {Hayes}, {Jordan}, {Kinney}, {Klaene}, {Kleinman},
  {Kron}, {Kresinski}, {Lee}, {Limmongkol}, {Lindenmeyer}, {Long}, {Loomis},
  {McGehee}, {Mantsch}, {Neilsen}, {Neswold}, {Newman}, {Nitta}, {Peoples},
  {Pier}, {Prieto}, {Prosapio}, {Rivetta}, {Schneider}, {Snedden}, \&
  {Wang}}]{Gunn:2006}
---. 2006, \aj, 131, 2332

\bibitem[{{Hodapp} {et~al.}(2004){Hodapp}, {Kaiser}, {Aussel}, {Burgett},
  {Chambers}, {Chun}, {Dombeck}, {Douglas}, {Hafner}, {Heasley}, {Hoblitt},
  {Hude}, {Isani}, {Jedicke}, {Jewitt}, {Laux}, {Luppino}, {Lupton}, {Maberry},
  {Magnier}, {Mannery}, {Monet}, {Morgan}, {Onaka}, {Price}, {Ryan},
  {Siegmund}, {Szapudi}, {Tonry}, {Wainscoat}, \& {Waterson}}]{Hodapp:2004}
{Hodapp}, K.~W., {et~al.} 2004, Astronomische Nachrichten, 325, 636

\bibitem[{{Hogg} {et~al.}(2001){Hogg}, {Finkbeiner}, {Schlegel}, \&
  {Gunn}}]{hoggpt:2001}
{Hogg}, D.~W., {Finkbeiner}, D.~P., {Schlegel}, D.~J., \& {Gunn}, J.~E. 2001,
  \aj, 122, 2129

\bibitem[{{Ivezi{\'c}} {et~al.}(2004){Ivezi{\'c}}, {Lupton}, {Schlegel},
  {Boroski}, {Adelman-McCarthy}, {Yanny}, {Kent}, {Stoughton}, {Finkbeiner},
  {Padmanabhan}, {Rockosi}, {Gunn}, {Knapp}, {Strauss}, {Richards},
  {Eisenstein}, {Nicinski}, {Kleinman}, {Krzesinski}, {Newman}, {Snedden},
  {Thakar}, {Szalay}, {Munn}, {Smith}, {Tucker}, \& {Lee}}]{Ivezic:2004}
{Ivezi{\'c}}, {\v Z}., {et~al.} 2004, Astronomische Nachrichten, 325, 583

\bibitem[{{Kaiser} {et~al.}(2010){Kaiser}, {Burgett}, {Chambers}, {Denneau},
  {Heasley}, {Jedicke}, {Magnier}, {Morgan}, {Onaka}, \& {Tonry}}]{Kaiser:2010}
{Kaiser}, N., {et~al.} 2010, in Society of Photo-Optical Instrumentation
  Engineers (SPIE) Conference Series, Vol. 7733, Society of Photo-Optical
  Instrumentation Engineers (SPIE) Conference Series

\bibitem[{{Lee} {et~al.}(2008){Lee}, {Beers}, {Sivarani}, {Allende Prieto},
  {Koesterke}, {Wilhelm}, {Re Fiorentin}, {Bailer-Jones}, {Norris}, {Rockosi},
  {Yanny}, {Newberg}, {Covey}, {Zhang}, \& {Luo}}]{Lee:2008}
{Lee}, Y.~S., {et~al.} 2008, \aj, 136, 2022

\bibitem[{{Lupton} {et~al.}(2001){Lupton}, {Gunn}, {Ivezi{\'c}}, {Knapp}, \&
  {Kent}}]{Lupton:2001}
{Lupton}, R., {Gunn}, J.~E., {Ivezi{\'c}}, Z., {Knapp}, G.~R., \& {Kent}, S.
  2001, in Astronomical Society of the Pacific Conference Series, Vol. 238,
  Astronomical Data Analysis Software and Systems X, ed. F.~R. {Harnden}, Jr.,
  F.~A. {Primini}, \& H.~E. {Payne}, 269

\bibitem[{{Magnier}(2006)}]{Magnier:2006}
{Magnier}, E. 2006, in The Advanced Maui Optical and Space Surveillance
  Technologies Conference

\bibitem[{{Magnier}(2007)}]{Magnier:2007}
{Magnier}, E. 2007, in Astronomical Society of the Pacific Conference Series,
  Vol. 364, The Future of Photometric, Spectrophotometric and Polarimetric
  Standardization, ed. C.~{Sterken}, 153

\bibitem[{{Magnier} {et~al.}(2008){Magnier}, {Liu}, {Monet}, \&
  {Chambers}}]{Magnier:2008}
{Magnier}, E.~A., {Liu}, M., {Monet}, D.~G., \& {Chambers}, K.~C. 2008, in IAU
  Symposium, Vol. 248, IAU Symposium, ed. W.~J. {Jin}, I.~{Platais}, \&
  M.~A.~C. {Perryman}, 553--559

\bibitem[{{Metcalfe} {et~al.}(2013){Metcalfe}, {Farrow}, {Cole}, {Draper},
  {Norberg}, {Burgett}, {Chambers}, {Denneau}, {Flewelling}, {Kaiser},
  {Kudritzki}, {Magnier}, {Morgan}, {Price}, {Sweeney}, {Tonry}, {Wainscoat},
  \& {Waters}}]{Metcalfe:2013}
{Metcalfe}, N., {et~al.} 2013, \mnras, 435, 1825

\bibitem[{{Oke} \& {Gunn}(1983)}]{Oke:1983}
{Oke}, J.~B., \& {Gunn}, J.~E. 1983, \apj, 266, 713

\bibitem[{{Onaka} {et~al.}(2008){Onaka}, {Tonry}, {Isani}, {Lee}, {Uyeshiro},
  {Rae}, {Robertson}, \& {Ching}}]{Onaka:2008}
{Onaka}, P., {Tonry}, J.~L., {Isani}, S., {Lee}, A., {Uyeshiro}, R., {Rae}, C.,
  {Robertson}, L., \& {Ching}, G. 2008, in Society of Photo-Optical
  Instrumentation Engineers (SPIE) Conference Series, Vol. 7014, Society of
  Photo-Optical Instrumentation Engineers (SPIE) Conference Series

\bibitem[{{Padmanabhan} {et~al.}(2008){Padmanabhan}, {Schlegel}, {Finkbeiner},
  {Barentine}, {Blanton}, {Brewington}, {Gunn}, {Harvanek}, {Hogg},
  {Ivezi{\'c}}, {Johnston}, {Kent}, {Kleinman}, {Knapp}, {Krzesinski}, {Long},
  {Neilsen}, {Nitta}, {Loomis}, {Lupton}, {Roweis}, {Snedden}, {Strauss}, \&
  {Tucker}}]{ubercal:2008}
{Padmanabhan}, N., {et~al.} 2008, \apj, 674, 1217

\bibitem[{{Pier} {et~al.}(2003){Pier}, {Munn}, {Hindsley}, {Hennessy}, {Kent},
  {Lupton}, \& {Ivezi{\'c}}}]{Pier:2003}
{Pier}, J.~R., {Munn}, J.~A., {Hindsley}, R.~B., {Hennessy}, G.~S., {Kent},
  S.~M., {Lupton}, R.~H., \& {Ivezi{\'c}}, {\v Z}. 2003, \aj, 125, 1559

\bibitem[{{Schlafly} \& {Finkbeiner}(2011)}]{Schlafly:2011}
{Schlafly}, E.~F., \& {Finkbeiner}, D.~P. 2011, \apj, 737, 103

\bibitem[{{Schlafly} {et~al.}(2010){Schlafly}, {Finkbeiner}, {Schlegel},
  {Juri{\'c}}, {Ivezi{\'c}}, {Gibson}, {Knapp}, \& {Weaver}}]{Schlafly:2010}
{Schlafly}, E.~F., {Finkbeiner}, D.~P., {Schlegel}, D.~J., {Juri{\'c}}, M.,
  {Ivezi{\'c}}, {\v Z}., {Gibson}, R.~R., {Knapp}, G.~R., \& {Weaver}, B.~A.
  2010, \apj, 725, 1175

\bibitem[{{Schlafly} {et~al.}(2012){Schlafly}, {Finkbeiner}, {Juri{\'c}},
  {Magnier}, {Burgett}, {Chambers}, {Grav}, {Hodapp}, {Kaiser}, {Kudritzki},
  {Martin}, {Morgan}, {Price}, {Rix}, {Stubbs}, {Tonry}, \&
  {Wainscoat}}]{Schlafly:2012}
{Schlafly}, E.~F., {et~al.} 2012, \apj, 756, 158

\bibitem[{{Schlegel} {et~al.}(1998){Schlegel}, {Finkbeiner}, \&
  {Davis}}]{SFD:1998}
{Schlegel}, D.~J., {Finkbeiner}, D.~P., \& {Davis}, M. 1998, \apj, 500, 525

\bibitem[{{Tonry} \& {Onaka}(2009)}]{Tonry:2009}
{Tonry}, J., \& {Onaka}, P. 2009, in Advanced Maui Optical and Space
  Surveillance Technologies Conference

\bibitem[{{Tonry} {et~al.}(2012{\natexlab{a}}){Tonry}, {Stubbs}, {Kilic},
  {Flewelling}, {Deacon}, {Chornock}, {Berger}, {Burgett}, {Chambers},
  {Kaiser}, {Kudritzki}, {Hodapp}, {Magnier}, {Morgan}, {Price}, \&
  {Wainscoat}}]{Tonry:2012md}
{Tonry}, J.~L., {et~al.} 2012{\natexlab{a}}, \apj, 745, 42

\bibitem[{{Tonry} {et~al.}(2012{\natexlab{b}}){Tonry}, {Stubbs}, {Lykke},
  {Doherty}, {Shivvers}, {Burgett}, {Chambers}, {Hodapp}, {Kaiser},
  {Kudritzki}, {Magnier}, {Morgan}, {Price}, \& {Wainscoat}}]{Tonry:2012}
---. 2012{\natexlab{b}}, \apj, 750, 99

\bibitem[{{Tucker} {et~al.}(2006){Tucker}, {Kent}, {Richmond}, {Annis},
  {Smith}, {Allam}, {Rodgers}, {Stute}, {Adelman-McCarthy}, {Brinkmann}, {Doi},
  {Finkbeiner}, {Fukugita}, {Goldston}, {Greenway}, {Gunn}, {Hendry}, {Hogg},
  {Ichikawa}, {Ivezi{\'c}}, {Knapp}, {Lampeitl}, {Lee}, {Lin}, {McKay},
  {Merrelli}, {Munn}, {Neilsen}, {Newberg}, {Richards}, {Schlegel},
  {Stoughton}, {Uomoto}, \& {Yanny}}]{Tucker:2006}
{Tucker}, D.~L., {et~al.} 2006, Astronomische Nachrichten, 327, 821

\bibitem[{{York} {et~al.}(2000){York}, {Adelman}, {Anderson}, {Anderson},
  {Annis}, {Bahcall}, {Bakken}, {Barkhouser}, {Bastian}, {Berman}, {Boroski},
  {Bracker}, {Briegel}, {Briggs}, {Brinkmann}, {Brunner}, {Burles}, {Carey},
  {Carr}, {Castander}, {Chen}, {Colestock}, {Connolly}, {Crocker}, {Csabai},
  {Czarapata}, {Davis}, {Doi}, {Dombeck}, {Eisenstein}, {Ellman}, {Elms},
  {Evans}, {Fan}, {Federwitz}, {Fiscelli}, {Friedman}, {Frieman}, {Fukugita},
  {Gillespie}, {Gunn}, {Gurbani}, {de Haas}, {Haldeman}, {Harris}, {Hayes},
  {Heckman}, {Hennessy}, {Hindsley}, {Holm}, {Holmgren}, {Huang}, {Hull},
  {Husby}, {Ichikawa}, {Ichikawa}, {Ivezi{\'c}}, {Kent}, {Kim}, {Kinney},
  {Klaene}, {Kleinman}, {Kleinman}, {Knapp}, {Korienek}, {Kron}, {Kunszt},
  {Lamb}, {Lee}, {Leger}, {Limmongkol}, {Lindenmeyer}, {Long}, {Loomis},
  {Loveday}, {Lucinio}, {Lupton}, {MacKinnon}, {Mannery}, {Mantsch}, {Margon},
  {McGehee}, {McKay}, {Meiksin}, {Merelli}, {Monet}, {Munn}, {Narayanan},
  {Nash}, {Neilsen}, {Neswold}, {Newberg}, {Nichol}, {Nicinski}, {Nonino},
  {Okada}, {Okamura}, {Ostriker}, {Owen}, {Pauls}, {Peoples}, {Peterson},
  {Petravick}, {Pier}, {Pope}, {Pordes}, {Prosapio}, {Rechenmacher}, {Quinn},
  {Richards}, {Richmond}, {Rivetta}, {Rockosi}, {Ruthmansdorfer}, {Sandford},
  {Schlegel}, {Schneider}, {Sekiguchi}, {Sergey}, {Shimasaku}, {Siegmund},
  {Smee}, {Smith}, {Snedden}, {Stone}, {Stoughton}, {Strauss}, {Stubbs},
  {SubbaRao}, {Szalay}, {Szapudi}, {Szokoly}, {Thakar}, {Tremonti}, {Tucker},
  {Uomoto}, {Vanden Berk}, {Vogeley}, {Waddell}, {Wang}, {Watanabe},
  {Weinberg}, {Yanny}, {Yasuda}, \& {SDSS Collaboration}}]{York:2000}
{York}, D.~G., {et~al.} 2000, \aj, 120, 1579

\bibitem[{{Zacharias} {et~al.}(2004){Zacharias}, {Urban}, {Zacharias},
  {Wycoff}, {Hall}, {Monet}, \& {Rafferty}}]{UCAC2:2004}
{Zacharias}, N., {Urban}, S.~E., {Zacharias}, M.~I., {Wycoff}, G.~L., {Hall},
  D.~M., {Monet}, D.~G., \& {Rafferty}, T.~J. 2004, \aj, 127, 3043

\end{thebibliography}

\end{document}